\documentclass[usegraphicx,usedcolumn,usenatbib]{mn2e}

\usepackage{color,epsfig,rotating,amssymb,longtable}
\usepackage{array,colortbl,lscape}
\def\HII{{H{\sc ii}}}

\usepackage{color}

\definecolor{dgreen}{rgb}{0,.5,.1} 
\definecolor{pink}{rgb}{.9,.2,.5}  
\definecolor{orange}{rgb}{.9,.4,0} 
\definecolor{darkred}{rgb}{.545,0.0,.0}
\begin{document}
\title[Dynamical masses of the CNSF clusters of NGC\,2903]
{On the derivation of dynamical masses of the stellar clusters in the circumnuclear region of NGC\,2903}
\author[G. F. H\"agele et al.]
{Guillermo F. H\"agele$^{1,2}$\thanks{E-mail: guille.hagele@uam.es},
\'Angeles I. D\'{\i}az$^{1}$,
 M\'onica V. Cardaci$^{1,3}$\thanks{PhD fellow of Ministerio de Educaci\'on y
    Ciencia, Spain}, 
  Elena Terlevich$^{4}$\thanks{Research
  Affiliate at IoA} \newauthor and Roberto Terlevich$^{4}$\footnotemark[3]
\\
$^{1}$ Departamento de F\'{\i}sica Te\'orica, C-XI, Universidad Aut\'onoma de
Madrid, 28049 Madrid, Spain\\ 
$^{2}$ Facultad de Cs Astron\'omicas y Geof\'isicas, Universidad Nacional de La
Plata, Paseo del Bosque s/n, 1900 La Plata, Argentina \\ 
$^{3}$ XMM Science Operations Centre, European Space Astronomy Centre of ESA,
P.O. Box 50727, 28080 Madrid, Spain\\ 
$^{4}$ INAOE, Tonantzintla, Apdo. Postal 51, 72000 Puebla, M\'exico\\ }

\maketitle
\begin{abstract}
Gas and star velocity dispersions have been derived for four circumnuclear 
star-forming regions (CNSFRs) and the nucleus of the spiral galaxy
NGC\,2903 using high resolution spectroscopy in the blue and far red. Stellar
velocity dispersions have been obtained from the CaII triplet (CaT) lines at
$\lambda\lambda$\,8494, 8542, 8662\,\AA, using cross-correlation techniques
while gas velocity dispersions have been measured by Gaussian fits to the
H$\beta$\,$\lambda$\,4861\,\AA\ line.  

The CNSFRs, with sizes of about 100 to 150\,pc in diameter, show a complex
structure at the {\it Hubble Space Telescope} (HST) resolution, with a good
number of subclusters with linear diameters between 3 and 8\,pc.   
Their stellar velocity dispersions range from 39 to 67 \,km\,s$^{-1}$. These
values, together with the sizes measured on archival HST images yield
{upper limits to the} dynamical masses for the individual star clusters
between 1.8 and 
8.7\,$\times$\,10$^6$\,M$_\odot$ and {upper limits to the} masses for the
whole CNSFR between 
4.9\,$\times$\,10$^6$ and 4.3\,$\times$\,10$^7$\,M$_\odot$.

The masses of the ionizing stellar population responsible for the \HII\ region
gaseous emission have been derived from their published H$\alpha$ luminosities
and are found to be between 1.9 and 2.5\,$\times$\,10$^6$\,M$_\odot$ for the
star-forming regions, and 2.1\,$\times$\,10$^5$\,M$_\odot$ for the galaxy
nucleus therefore constituting between 1 and 4 per cent of the total dynamical
mass. 

In the CNSFR star and gas velocity dispersions are found to differ by about 
20\,km\,s$^{-1}$ with the H$\beta$ lines being narrower than both the stellar
lines and the [O{\sc iii}]\,$\lambda$\,5007\,\AA\ lines. The ionized gas
kinematics is complex; two different kinematical components seem to be present
as evidenced by different widths and Doppler shifts. 

The line profiles in the spectra of the galaxy nucleus, however, are
consistent with the presence of a single component with radial velocity and
velocity dispersion close to those measured for the stellar absorption lines. 

The presence and reach of two distinct components in the emission lines in
ionized regions and the influence that this fact could have on the observed
line ratios is of major interest for several reasons, among others, 
the classification of the activity in the central
regions of galaxies, the inferences about the nature of the
source of ionization for the two components and the derivation of the gaseous
chemical abundances.

\end{abstract}


\section{Introduction}

The inner ($\sim$1\,Kpc) parts of some spiral galaxies show high star
formation rates and this star formation is frequently arranged in a ring or
pseudo-ring pattern around their nuclei. This fact seems to correlate with the
presence of bars
\citep{1979ApJ...233...67R,1995A&A...301..649F,1999ApJ...525..691S,2005ApJ...632..217S,2005ApJ...630..837J}
and, in fact, computer models which simulate the behaviour of gas in galactic
potentials have shown that nuclear rings may appear as a consequence 
of matter infall owing to resonances present at the bar edges
\citep{1985A&A...150..327C,1992MNRAS.259..328A}.

{These circumnuclear star-forming regions (CNSFRs) are expected to be amongst
the highest metallicity regions as 
corresponds to their position near the galactic center. This fact makes 
the analysis of these regions complicated since, in general,
their low excitation makes any temperature sensitive line too weak to be
measured \citep{2007MNRAS.382..251D}, particularly against a strong underlying
stellar continuum. As was 
pointed out in H\"agele et al.\ (2007; hereafter Paper I)
\nocite{2007MNRAS.378..163H} the [O{\sc iii}] emission lines in CNSFRs are
generally very weak, 
and in some cases unobservable. The low 
value of these collisionally excited lines can be explained by their over-solar
metal abundances \citep[e.g.\ ][]{1993A&A...277..397B}. The equivalent width
of the emission lines are lower than those shown by the disc \HII\ regions
\citep[see for
  example][]{1989AJ.....97.1022K,1997AJ....113..975B,1999ApJ...510..104B}. 
Combining GEMINI data 
and a grid of photo-ionization models
\cite{2008A&A...482...59D} conclude that the contamination of the continua
of CNSFRs by underlying contributions from both old bulge stars and stars
formed in the ring in previous episodes of star formation (10-20\,Myr) yield
the observed low equivalent widths. 

CNSFRs, with sizes going from a few tens to a few hundreds of parsecs 
\citep[e.g.][]{2000MNRAS.311..120D} seem to be made of several \HII\ regions
ionized by luminous compact stellar clusters whose sizes, as measured from
high spatial resolution {\it Hubble Space Telescope} (HST) images, are seen to
be of only a few parsecs. 
%
%
Although these \HII\ regions are very luminous (M$_v$ between -12 and -17) not
much is known about their  
kinematics or dynamics for both the ionized gas and the stars. It could be
said that the worst known parameter of these ionizing clusters is their mass. 
There are different methods to estimate the mass of a stellar
cluster. 
As derived with the use of population synthesis models their masses suggest
that these  clusters are gravitationally bound and that they might evolve into
globular cluster configurations \citep{1996AJ....111.2248M}. 
Classically one assumes that the system is virialized and determines
the total mass inside a radius by applying the virial theorem to the observed 
velocity dispersion of the stars ($\sigma_{\ast}$). The stellar velocity
dispersion is however hard to measure in young stellar clusters (a few
million-years old) due to the 
shortage of prominent stellar absorption lines. The optical continuum between
3500 and 7000\,\AA\ shows very few lines since the light at  these wavelengths
is dominated by OB stars which have weak absorption lines at the same
wavelengths of the nebular emission lines (Balmer H and He{\sc i} lines). As
pointed out by several authors \citep[e.g.\ ][]{1996ApJ...466L..83H}, at
longer  
wavelengths (\,8500\,\AA) the contamination due to nebular lines is much
smaller  and since red supergiant stars, if present, dominate the
near-infrared (IR)
light where the Ca{\sc ii} $\lambda\lambda$\,8498, 8542, 8662\,\AA\ triplet
(CaT) lines are found, these should be easily observable allowing the
determination of $\sigma_{\ast}$
\citep{1990MNRAS.242P..48T,1994A&A...288..396P}. 
}

NGC\,2903 (UGC\,5079) is a well studied galaxy classified as a SB(rs)bc by
\cite{1991trcb.book.....D}. Its coordinates are $\alpha_{\rm
2000}$=09$^h$\,32$^m$\,10\fs1, $\delta_{\rm
2000}$=+21$^{\circ}$\,30\arcmin\,03\arcsec \citep{1991trcb.book.....D}.
These authors derived a B$_{T}$ (mag) equal to 9.7 and an E(B-V)$_{gal}$ (mag)
of 0.031. We adopt the distance derived by \cite{1984AAS...56..381B} which is
equal to 8.6\,Mpc, giving a linear scale of $\sim$\,42\,pc\,arcsec$^{-1}$.

The Paschen $\alpha$ image obtained with the Hubble Space Telescope (HST)
reveals in NGC\,2903 the presence of a nuclear ring-like 
morphology with an apparent diameter of approximately 15\,\arcsec\,=\,625\,pc
\citep{2001MNRAS.322..757A}. This structure is also seen, though less
prominent, in the H$\alpha$ observations from \cite{1997A&A...325...81P}. A
large number of stellar clusters are identified 
on  high resolution infrared images in the K' and H bands, which do not
coincide spatially with the bright \HII\ regions. A possible interpretation of
this is that the stellar clusters are the result of the evolution of giant
\HII\ regions \citep[e.g.][]{2001MNRAS.322..757A}. The global star
formation rates in the nuclear ring, as derived from  its H$\alpha$ luminosity
is found to be 0.1\,M$_\odot$\,yr$^{-1}$ by \cite{1997A&A...325...81P} and
0.7\,M$_\odot$\,yr$^{-1}$ by \cite{2001MNRAS.322..757A}. From CO emission 
observations, Planesas et al.\ derive a mass of molecular gas (H$_2$) of
1.8\,$\times$\,10$^{8}$\,M$_\odot$ inside a circle 1\,Kpc in diameter.

This is the second paper of a series to study the peculiar conditions of
star formation in circumnuclear regions of early type spiral galaxies, 
in particular the kinematics of the connected stars and gas.
In this paper we present high-resolution  far-red spectra
and  stellar velocity dispersion
measurements ($\sigma_{\ast}$) along the line of sight for four
CNSFRs and the nucleus of the spiral galaxy NGC\,2903. We have also measured
the ionized gas velocity dispersions ($\sigma_{g}$) from high-resolution blue
spectra
using Balmer H$\beta$ and [O{\sc iii}] emission
lines. The comparison between $\sigma_{\ast}$ and $\sigma_{g}$  might throw
some light on the yet unsolved issue about the validity of the gravitational
hypothesis for the origin of the supersonic motions observed in the ionized
gas in Giant \HII\ regions \citep*{1999MNRAS.302..677M}.

In Section 2 we describe the observations and data reduction. We present the 
results in Section 3, the dynamical mass derivation in Section 4 and the
ionizing star cluster properties in Section 5. We discuss all our results in
Section 6. Finally, summary and conclusions are given in Section 7.

\section{Observations and data reduction}
\label{Obs}




\begin{table*}
\centering
\caption[]{Journal of Observations}
\begin{tabular} {l c c c c c c c}
\hline
 Date & Spectral range &       Disp.          & R$^a_{\rm{FWHM}}$  & Spatial
 res.            & PA   & Exposure Time & {seeing$_{\rm{FWHM}}$} \\
         &     (\AA)          & (\AA\,px$^{-1}$) &    & (\arcsec\,px$^{-1}$) &
 ($ ^{o} $) & (sec) & (\arcsec)  \\
\hline
2000 February 4 & 4779-5199  &  0.21  &  $\sim$\,12500  &  0.38  &    50    &  3\,$\times$\,1200 & {1.2}\\
                & 8363-8763  &  0.39  &  $\sim$\,12200  &  0.36  &    50    &  3\,$\times$\,1200 &  \\
2000 February 5 & 4779-5199  &  0.21  &  $\sim$\,12500  &  0.38  &   345    &  3\,$\times$\,1200 & {1.6}\\
                & 8363-8763  &  0.39  &  $\sim$\,12200  &  0.36  &   345    &  3\,$\times$\,1200 & \\
\hline
\multicolumn{7}{l}{$^a$R$_{\rm{FWHM}}$\,=\,$\lambda$/$\Delta\lambda_{\rm{FWHM}}$}
\end{tabular}
\label{journal}
\end{table*}


The data were acquired in 2000 February using the two arms of the ISIS
spectrograph attached to the 4.2-m William Herschel Telescope (WHT) of the
Isaac Newton Group (ING) at the Roque de los Muchachos Observatory on the
Spanish island of La Palma. {The CCD
detectors EEV12 and TEK4 were used for the blue and red arms with a factor of
2 binning in both the ``x" and ``y" directions with resultant spatial
resolutions of 0.38 and 0.36 \,arcsec\,pixel$^{-1}$ for the blue and red
configurations respectively. The H2400B and R1200R gratings were used to cover
the wavelength ranges from 4779 to 5199\,\AA\ ($\lambda_c$\,=\,4989\,\AA) in
the blue and  from 8363 to 8763\,\AA\ ($\lambda_c$\,=\,8563\,\AA) in the red
with resultant spectral dispersions  of 0.21 and 0.39 \AA\ per pixel
respectively, providing a comparable velocity resolution of about 13 km s$
^{-1}$.} A slit width of 1\,arcsec was used which, 
combined with the spectral dispersions, yielded spectral resolutions of about
0.4 and 0.7\,\AA\ FWHM in the blue and the red, respectively, measured on the
sky lines. Table \ref{journal} summarizes the instrumental configuration and
observation details. Several bias and sky flat field frames were taken at the
beginning and end of each night in both arms. In addition, two lamp flat field
and one calibration lamp exposures per telescope position were performed. The
calibration lamp used was CuNe+CuAr.

\label{Data reduction}
The data were processed and analysed using IRAF\footnote{IRAF: the Image
Reduction and Analysis Facility is distributed by the National Optical
Astronomy Observatories, which is operated by the Association of Universities
for Research in Astronomy, Inc. (AURA) under cooperative agreement with the
National Science Foundation (NSF).} routines including the usual procedures of
removal of cosmic rays, bias subtraction, division by a normalized flat field
and wavelength calibration. Further details concerning each of these steps can
be found in Paper I.




















Neither atmospheric extinction nor flux calibration were performed, since our
purpose was to measure radial velocities and velocity dispersions. With this
purpose, spectra of 11 template velocity stars were  acquired to provide good
stellar reference frames in the same system as the galaxy spectra for the
kinematic analysis in the far-red. They correspond to late-type giant and
supergiant stars which have strong CaT features
\citep[see][]{1989MNRAS.239..325D}. The spectral types, luminosity classes and
dates of observation of the stellar reference frames used as templates are
listed in Table 2 of Paper I.

\label{Results}
\section{Results}

\begin{figure*}
    \centering
    \includegraphics[width=0.48\textwidth]{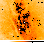}
\hspace{0.2cm}
\includegraphics[width=0.48\textwidth]{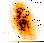}
\caption[]{Left: F606W (wide V) image centred on NGC\,2903 obtained with the
  WFPC2 camera (PC1) of the HST. Right: HST-NICMOS (NIC3) image obtained
  through the F160W filter. For both images the orientation is north up, east
  to the left. The location and P.A. of the WHT-ISIS slit positions, together
  with identifications of the CNSFRs extracted, are marked.}
\label{hst-slits}
\end{figure*}

Two different slit positions (S1 and S2) were chosen in order 
to observe the nucleus of the galaxy and  4 CNSFRs. 
Fig.\ \ref{hst-slits} shows them, superimposed on photometrically calibrated 
optical and infrared 
images of the circumnuclear region of NGC\,2903 acquired with
the  Wide Field and Planetary Camera 2 (WFPC2; PC1) and the Near-Infrared
Camera and Multi-Object Spectrometer (NICMOS) Camera 3 (NIC3) on board  the 
HST.
These images have been downloaded  from the Multimission Archive at STScI 
(MAST)\footnote{http://archive.stsci.edu/hst/wfpc2}. The optical image
was obtained through the F606W (wide V) filter, and the near-IR one,
through the F160W (H). The CNSFRs have been labelled following 
the same nomenclature as in \cite{1997A&A...325...81P} from
their identifications on  H$\alpha$ maps. The plus symbol in both
panels of Fig.\ \ref{hst-slits} represents the position of the nucleus as given by
the Two Micron All Sky Survey Team, 2MASS Extended Ojects - Final Release
\citep{2003AJ....125..525J}.

\begin{figure*}
\includegraphics[width=.41\textwidth,angle=0]{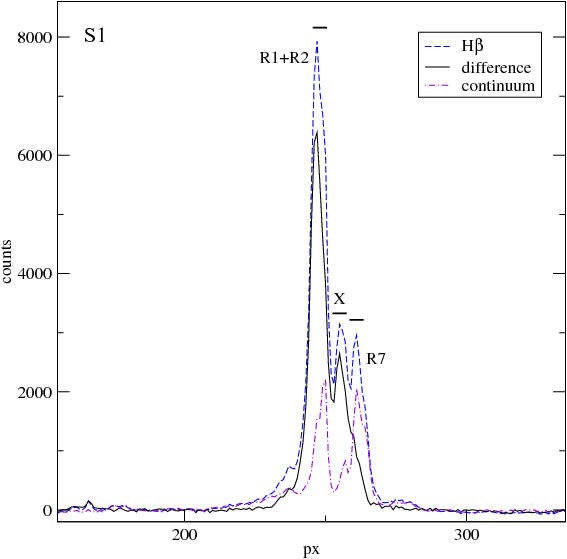}
\hspace{0.2cm}
\includegraphics[width=.41\textwidth,angle=0]{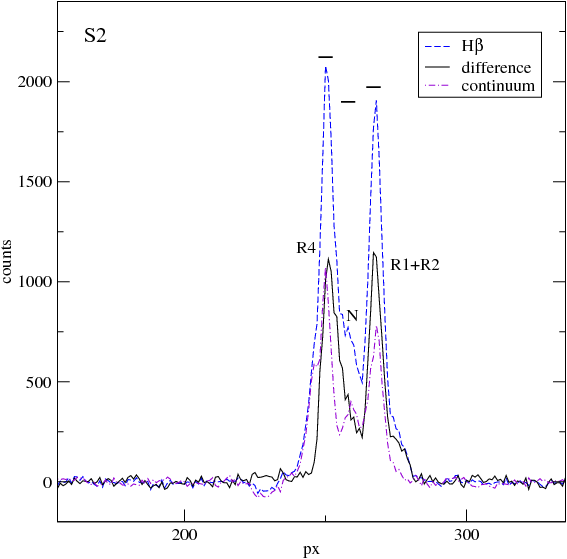}\\
\vspace{0.3cm}
\includegraphics[width=.41\textwidth,angle=0]{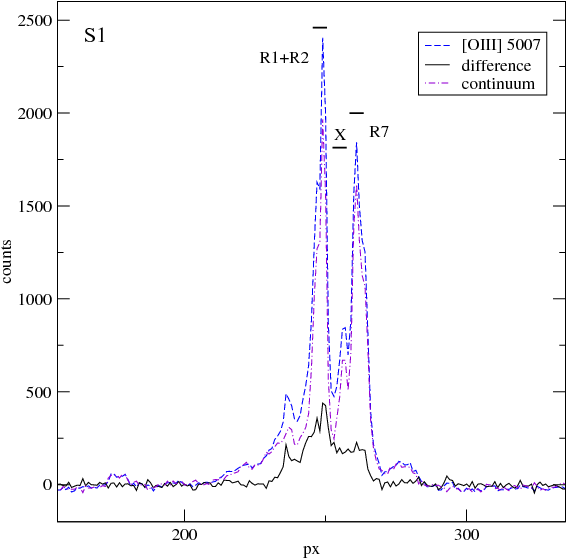}
\hspace{0.2cm}
\includegraphics[width=.41\textwidth,angle=0]{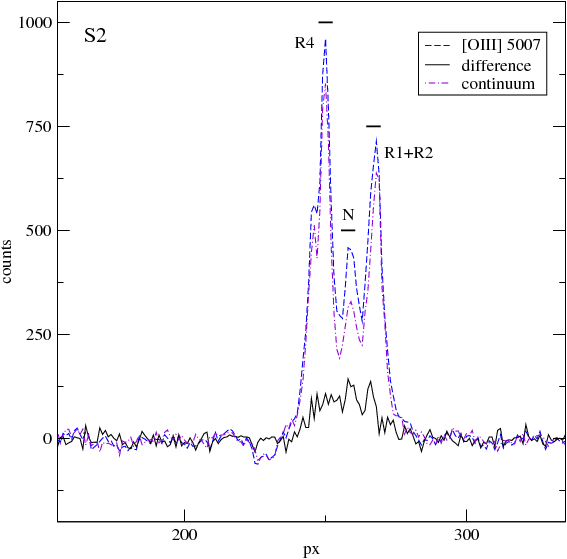}\\
\vspace{0.3cm}
\includegraphics[width=.41\textwidth,angle=0]{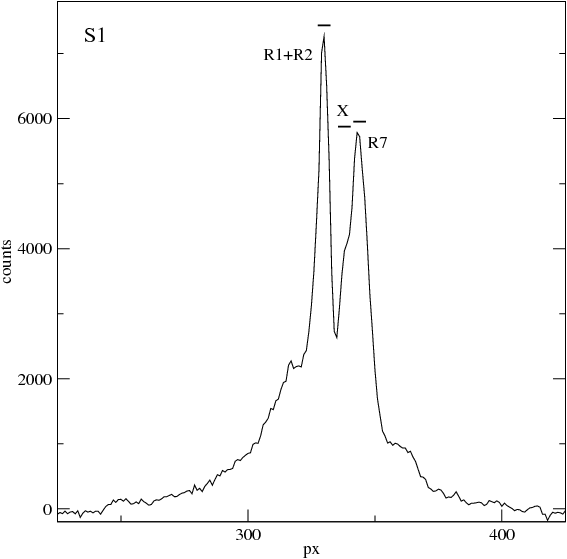}
\hspace{0.2cm}
\includegraphics[width=.41\textwidth,angle=0]{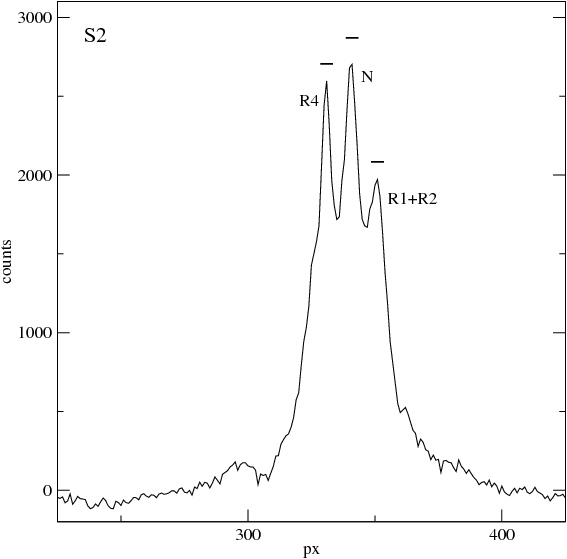}\\
\caption{Spatial profiles of H$\beta$, [O{\sc iii}]\,$\lambda$\,5007\AA\ and
  the far red (upper, middle and lower panels respectively) for each
  slit. For the emission lines, the profiles correspond to 
  line+continuum (dashed line), continuum (dashed-dotted line) and the 
  difference between them (solid line), representing the pure emission from
  H$\beta$ and [O{\sc iii}] respectively. For the far red profiles, the solid
  lines represent the continuum. Pixel number increases to the
  North. Horizontal small lines show the location of the CNSFRs and nuclear
  apertures.} 
\label{profiles}
\end{figure*}

Fig.\ \ref{profiles} shows the spatial profiles in the H$\beta$ and  [O{\sc
iii}]\,5007\,\AA\ emission lines (upper and middle panels) and the far-red 
continuum (lower panels) along each slit position. The emission line
profiles have been generated by collapsing 11 pixels of the spectra in the
direction of the resolution at the central position of the lines in the rest
frame, $\lambda$\,4861 and $\lambda$\,5007\AA\ respectively, and are plotted
as dashed lines. Continuum profiles were generated by collapsing 11 resolution
pixels centred at 11\,\AA\ to the blue of each emission line and are
plotted as dash-dotted lines. The difference between the two, shown by solid
lines, corresponds to the pure emission. The far-red 
continuum has been generated by collapsing 11 pixels centred at
$\lambda$\,8620\AA.

The regions of the frames to be extracted into one-dimensional spectra 
corresponding to each of the identified CNSFRs, were selected on these 
profiles with reference to the continuum emission both in the blue and in the 
red. These regions are marked by horizontal lines and labelled in the
figure. In the H$\beta$ profile of slit position S1, an almost pure emission
knot between regions R1+R2 and R7 can be seen with a very weak continuum. We
labelled this region as X. In
the case of the [O{\sc iii}]\,5007\,\AA\ emission line profile, the main
contributor comes from the continuum since the actual emission line is very
weak as expected for high metallicity \HII\ regions
\citep{2007MNRAS.382..251D}. Spectra in slit position S1 are placed along the
circumnuclear regions located to the North and North-East of the nucleus and
therefore any contribution from the underlying galaxy bulge is difficult to
assess. In the case of slit position S2 which crosses the galactic nucleus,
this fact can be used to estimate that contribution. For  the blue spectra the
light from the underlying bulge is almost negligible amounting to, at most, 5
per cent at the H$\beta$ line. For the red spectra, the bulge contribution is
more important. From Gaussian fits to the $\lambda$\,8620\,\AA\ continuum
profile we find that it amounts to about 15 per cent for the lowest surface
brightness region, R1+R2. On the other hand, the analysis of the broad near-IR
HST-NICMOS image shown in Fig.\ \ref{hst-slits}, indicates less contrast between
the emission from the regions and the underlying bulge which is very close to
the image background emission. Its contribution is about 25 per cent for the
weak regions, R2 and R7 along position slit S1.

\begin{figure*}
\includegraphics[width=.48\textwidth,height=.30\textwidth]{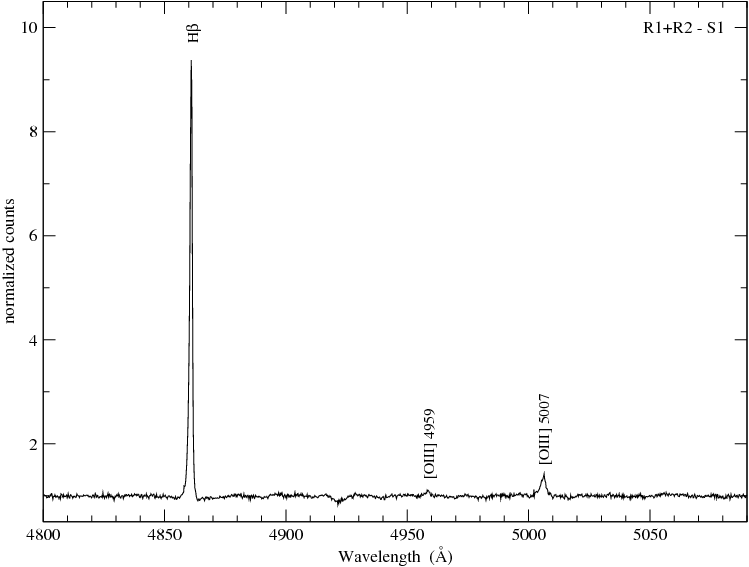}
\includegraphics[width=.48\textwidth,height=.30\textwidth]{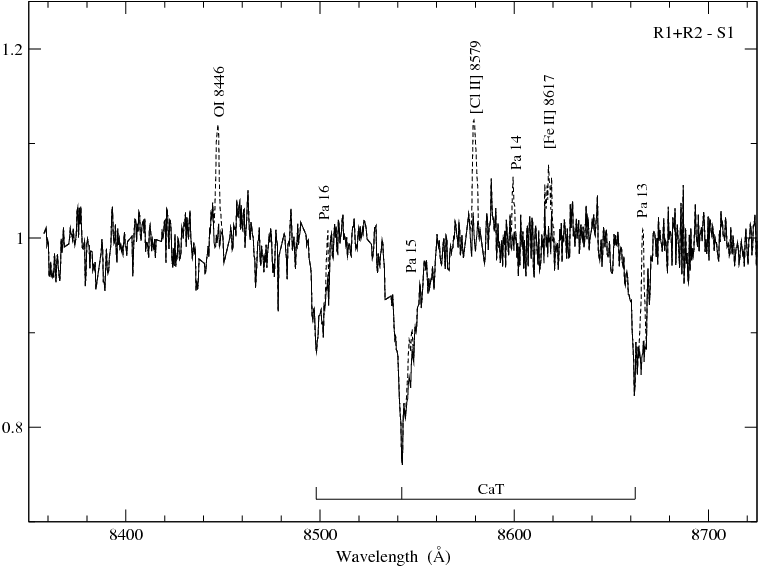}\\
\vspace{0.2cm}
\includegraphics[width=.48\textwidth,height=.30\textwidth]{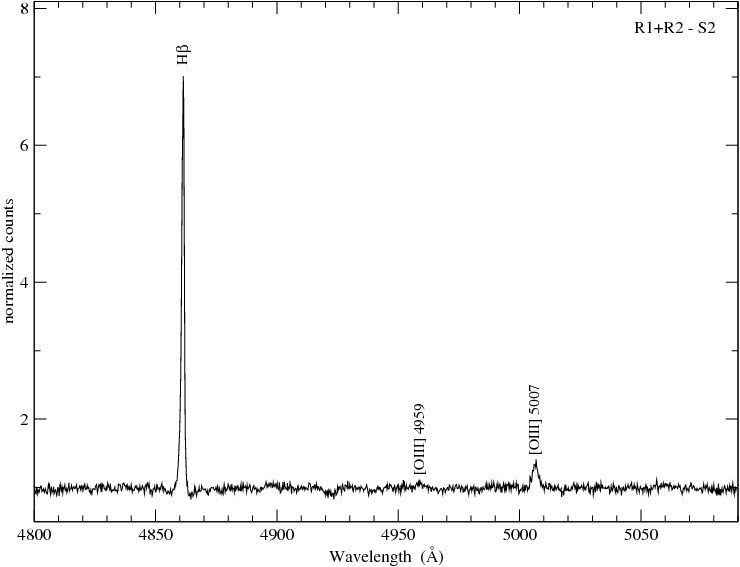}
\includegraphics[width=.48\textwidth,height=.30\textwidth]{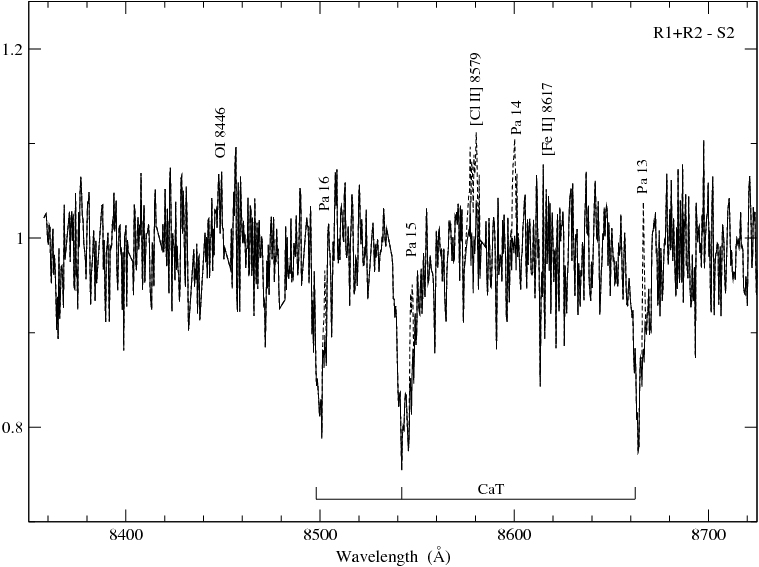}\\
\vspace{0.2cm}
\includegraphics[width=.48\textwidth,height=.30\textwidth]{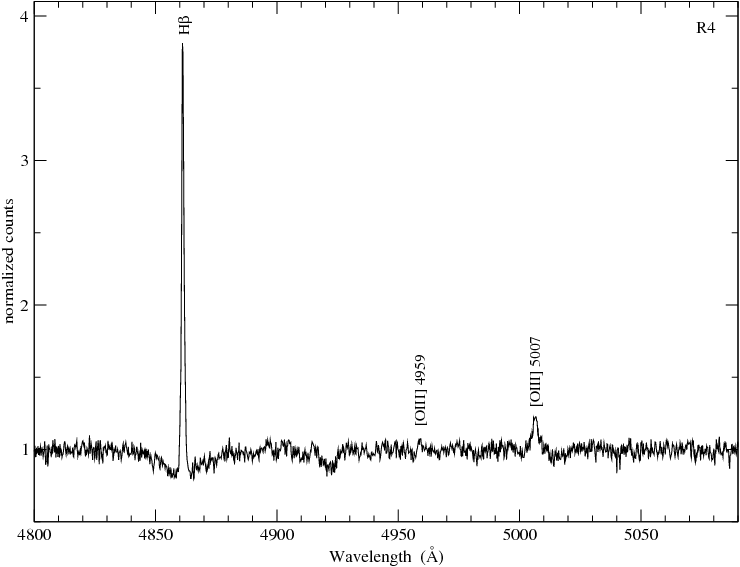}
\includegraphics[width=.48\textwidth,height=.30\textwidth]{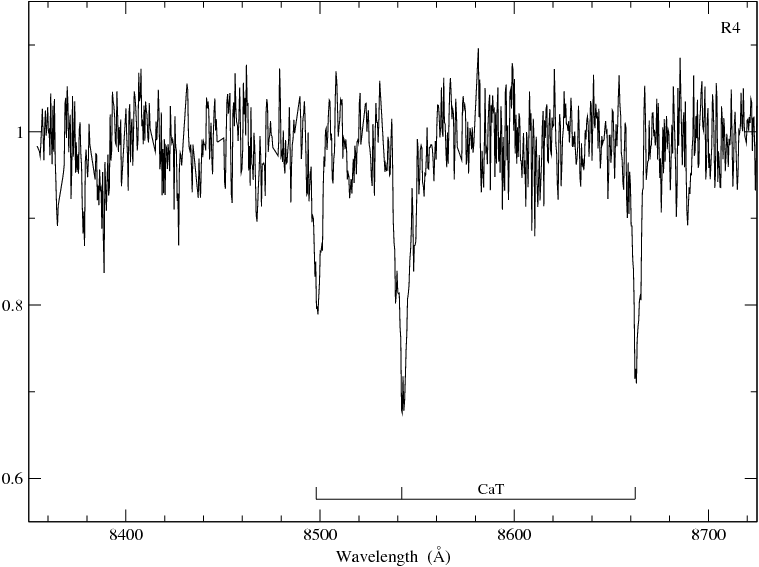}\\
\vspace{0.2cm}
\includegraphics[width=.48\textwidth,height=.30\textwidth]{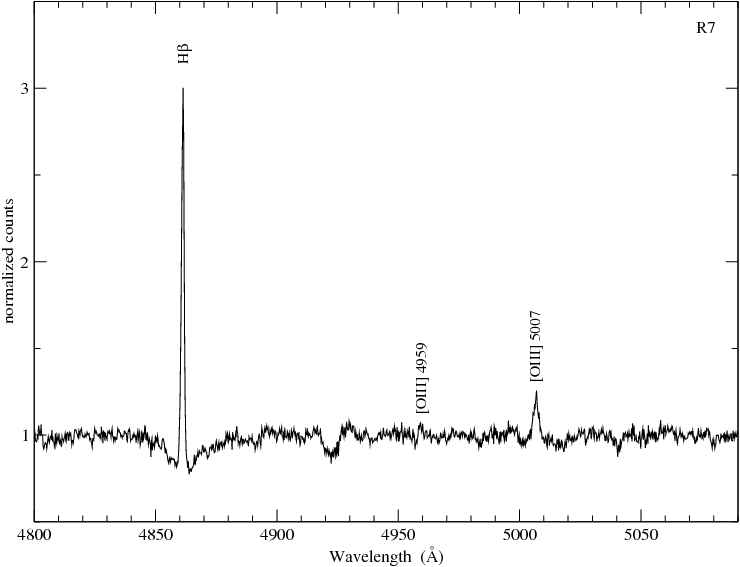}
\includegraphics[width=.48\textwidth,height=.30\textwidth]{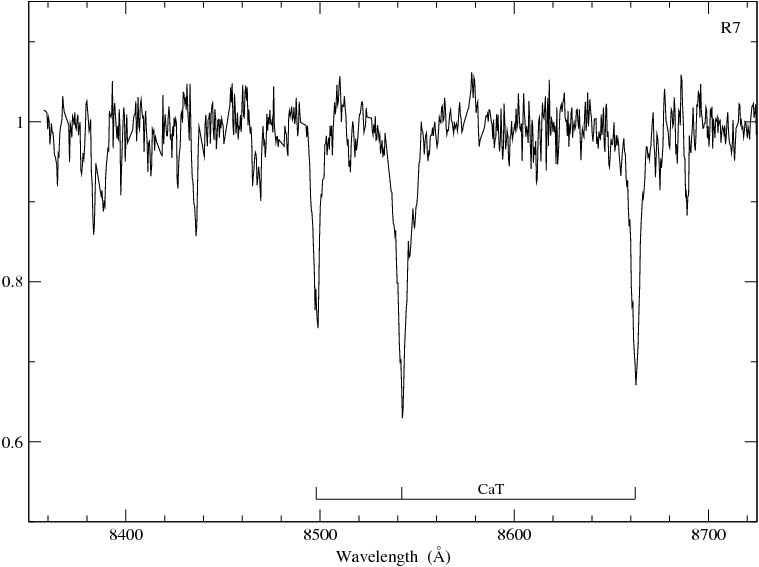}
\caption{Blue (left) and red (right) rest frame normalized spectra of the
  observed CNSFRs. For R1+R2, the dashed line shows the obtained spectrum;
  the solid line represents the spectrum after subtracting the emission lines
  (see text). Notice the absence of conspicuous emission lines in the  
  red spectral range for R4 and R7.} 
\label{spectra}
\end{figure*}

\begin{figure*}
\hspace{0.4cm}
\includegraphics[width=.48\textwidth,height=.30\textwidth,angle=0]{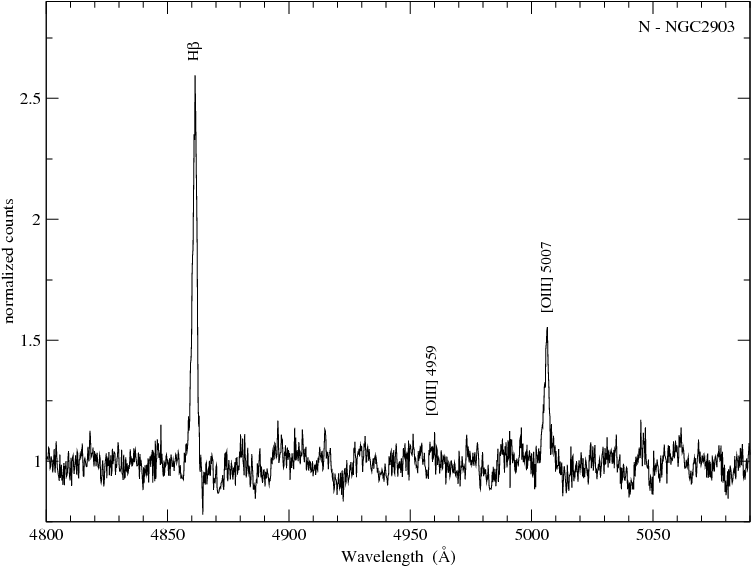}
\includegraphics[width=.48\textwidth,height=.30\textwidth,angle=0]{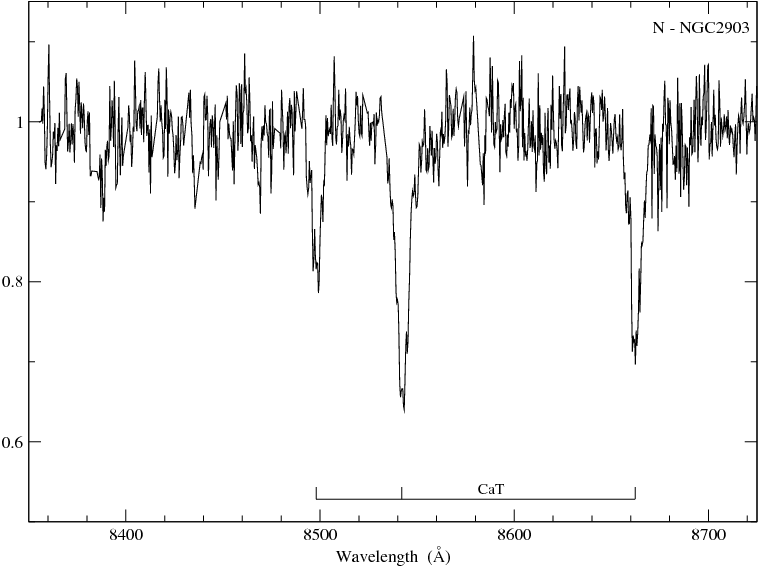}
\caption{Blue (left) and red (right) rest frame normalized spectra of the nucleus.}
\label{spectrum-N}
\end{figure*}

Figs.\ \ref{spectra} and \ref{spectrum-N} show the spectra of the observed
regions and the nucleus, respectively, split into two panels corresponding to
the blue and the red spectral range. The blue spectra show the Balmer H$\beta$
recombination line and the weak collisionally excited [O{\sc iii}] lines at
$\lambda\lambda$\,4959,\,5007\,\AA. 
The red spectra show the stellar CaII triplet  lines
in absorption (CaT) at $\lambda\lambda$ 8498, 8542, 8662\,\AA. 
In some cases, as for example R1+R2,  these 
lines are contaminated by Paschen emission which occurs at wavelengths very 
close to those of the CaT lines. Other emission features, such as O{\sc
i}\,$\lambda$\,8446, [Cl{\sc ii}]\,$\lambda$\,8579, Pa\,14 and [Fe{\sc
ii}]\,$\lambda$\,8617 are also present. In this case, a 
single Gaussian fit to the emission lines was performed and the lines were
subsequently subtracted after checking that the theoretically expected ratio
between the Paschen lines was satisfied. The observed
red spectra of R1+R2 are plotted in Fig.\ \ref{spectra} with a dashed line.  
The solid lines show the subtracted spectra.

\begin{figure*}
\hspace{0.0cm}
\vspace{0.2cm}
\includegraphics[width=.48\textwidth,height=.30\textwidth]{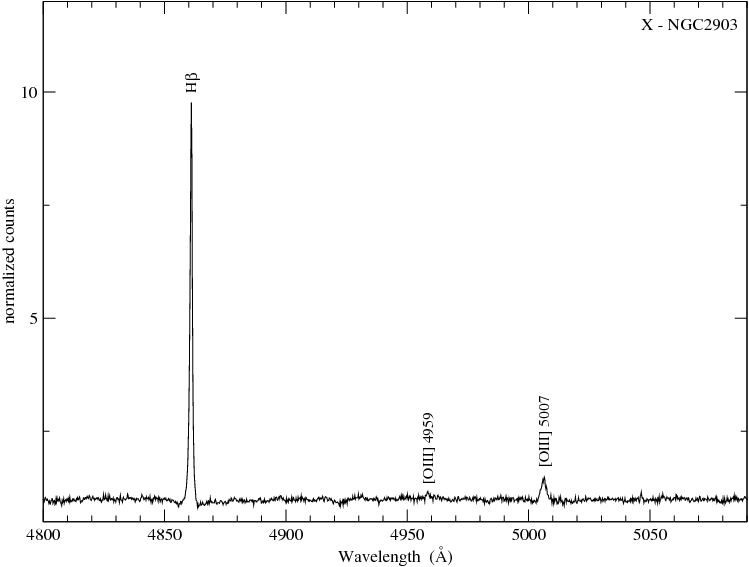}
\includegraphics[width=.48\textwidth,height=.30\textwidth]{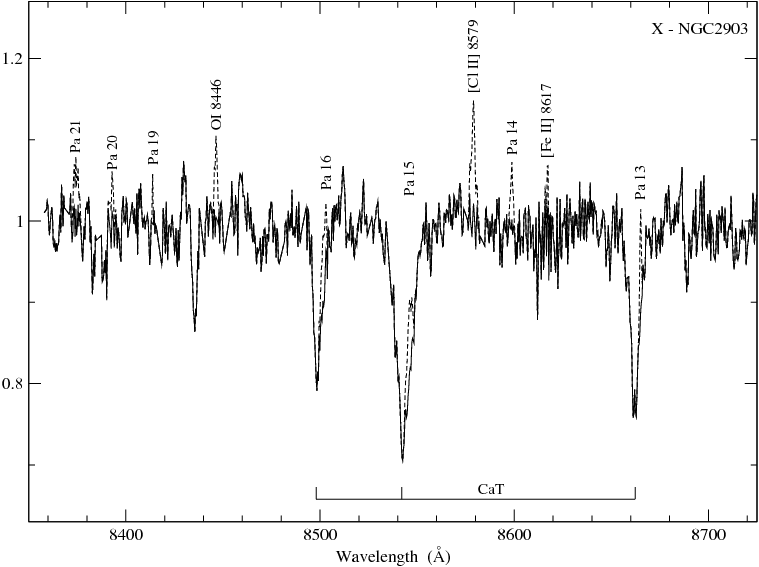}\\
\caption[]{Idem as Fig.\ \ref{spectra} for region X. The dashed line shows the obtained
  spectrum; the solid line represents the spectrum after subtracting the
  emission lines (see text).} 
\label{spectraemiknot}
\end{figure*}

Fig.\ \ref{spectraemiknot} shows the spectrum of the almost pure emission
knot marked in the profiles of NGC\,2903, region X. In this case the blue
range of the spectrum presents very intense emission lines, while in the red
range we can appreciate a spectrum similar to those shown by the CNSFRs
studied by Planesas et al.\ (1997) suggesting a region with similar characteristics but with low continuum surface brightness.

\subsection{Kinematics of stars and ionized gas} 
\label{Method}

A detailed description of the methods and techniques used to derive the values
of radial velocities and velocity dispersions as well as sizes, masses and
emission line fluxes has been given in Paper I. Therefore only a brief summary
is given below.

\subsubsection*{Stellar analysis}

Stellar radial velocities and velocity dispersions were obtained from the
conspicuous CaT absorption lines using the cross-correlation technique,
described in detail by \cite{1979AJ.....84.1511T}. This method requires the
comparison with a stellar template that represents the stellar population that
best reproduces the absorption features. This has been built from a set of 11
late-type giant and supergiant stars with strong CaT absorption lines. 
{We have followed the work by \cite{1995ApJS...99...67N}
with the variation introduced by \cite{1997A&A...323..749P} of using the
individual stellar templates instead of an average. This procedure will allow
us to correct for the known possible mismatches between template stars and the
region composite spectrum.} The
implementation of the method in the external package of IRAF XCSAO
\citep{1998PASP..110..934K} has been used, as explained in Paper I.


To determine the line-of-sight stellar velocity and velocity dispersion 
along each slit,  extractions 
were made every two pixels for slit position S1 and every three pixels for
slit position S2, with one pixel overlap between consecutive 
extractions in this latter case. In this way the S/N ratio and the spatial
resolution were optimised. The stellar velocity dispersion was
estimated at the position of each CNSFR and the nucleus using an aperture 
of five pixels in all cases, which corresponds to
1.0\,$\times$\,1.8\,arcsec$^2$. The velocity dispersion ($\sigma$) of the
stars ($\sigma_{\ast}$) is taken as the average of the $\sigma$ values found for each
stellar template, and its error is taken as the dispersion of the individual
values of $\sigma$ and the rms of the residuals of the wavelength fit. These
values  are listed in column 3 of Table \ref{disp} along with their
corresponding errors.


\begin{table*}
{\footnotesize
\centering
\caption[]{Velocity dispersions.}
\begin{center}
\begin{tabular} {@{}l c c c c c c c c c@{}}
\hline
        &      &                  &  \multicolumn{2}{c}{{\it 1 component}} &
        \multicolumn{4}{c}{{\it 2 components}}    \\
        &      &           &       &      &  \multicolumn{2}{c}{{\it narrow}}
        & \multicolumn{2}{c}{{\it broad}} &   \\
 Region & Slit & $\sigma_{\ast}$  & $\sigma_{gas}$(H$\beta$) &
 $\sigma_{gas}$([O{\sc iii}]) &  $\sigma_{gas}$(H$\beta$)   &
        $\sigma_{gas}$([O{\sc iii}]) & $\sigma_{gas}$(H$\beta$)  &
        $\sigma_{gas}$([O{\sc iii}]) &  $\Delta$v$_{nb}$  \\
\hline
R1+R2 & S1 & 60$\pm$3 & 34$\pm$2 & 73$\pm$8  & 23$\pm$2 & 26$\pm$8 & 51$\pm$3  & 93$\pm$9   & 30   \\
R1+R2 & S2 & 64$\pm$3 & 35$\pm$2 & 71$\pm$9  & 27$\pm$2 & 27$\pm$7 & 53$\pm$4  & 95$\pm$10  & 35   \\
R4    & S2 & 44$\pm$3 & 32$\pm$2 & 76$\pm$10 & 20$\pm$2 & 35$\pm$9 & 47$\pm$4  & 89$\pm$8   & -10  \\
R7    & S1 & 37$\pm$3 & 32$\pm$4 & 59$\pm$10 & 29$\pm$5 & 17$\pm$5 & 34$\pm$8  & 67$\pm$8   & 35   \\[2pt]
N     & S2 & 65$\pm$3 & 59$\pm$3 & 59$\pm$7  & 47$\pm$4 & 27$\pm$7 & 99$\pm$13 & 83$\pm$12  & 20   \\[2pt]
X     & S1 & 38$\pm$3 & 32$\pm$2 & 66$\pm$7  & 22$\pm$2 & 29$\pm$5 & 54$\pm$4  & 75$\pm$8   & 15  \\
\hline
\multicolumn{9}{@{}l}{velocity dispersions in km\,s$^{-1}$} \\
\end{tabular}
\label{disp}
\end{center}}
\end{table*}


The radial velocities have been determined directly
from the position of the main peak of the cross-correlation of each galaxy
spectrum with each template in the rest frame. The average of these values is
the final adopted radial velocity.

\subsubsection*{Ionized gas analysis}

The velocity dispersion of the ionized gas was estimated for each observed 
CNSFR and for the galaxy nucleus from  gaussian fits to the H$\beta$ 
and [O{\sc iii}]\,$\lambda$\,5007\,\AA\ emission lines using five pixel 
apertures, corresponding to 1.0\,$\times$\,1.9\,arcsec$^2$. For a single
Gaussian fit,  the position and width of a given emission line is taken as the
average of the fitted Gaussians to the whole line using three different
suitable continua \citep{2000MNRAS.317..907J}, and their errors are given by  
the dispersion of these measurements taking into account the rms of the
wavelength calibration.

\begin{figure*}
\includegraphics[width=.48\textwidth,height=.30\textwidth,angle=0]{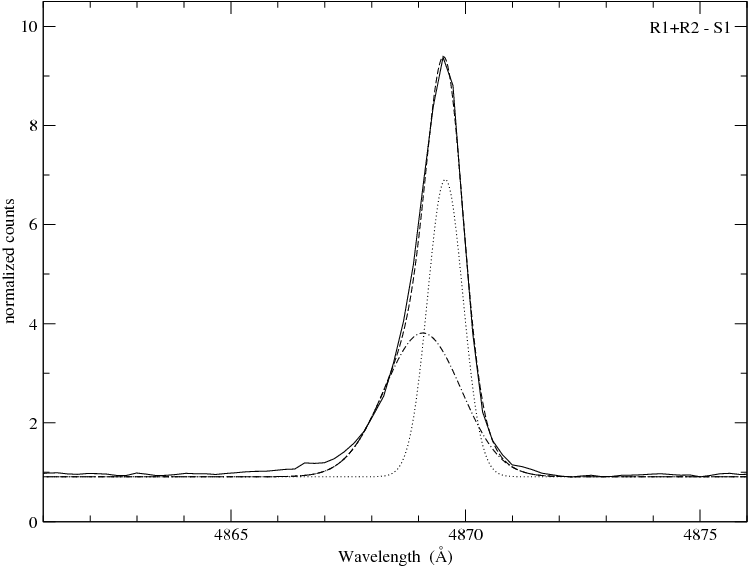}\hspace*{0.2cm}
\includegraphics[width=.48\textwidth,height=.30\textwidth,angle=0]{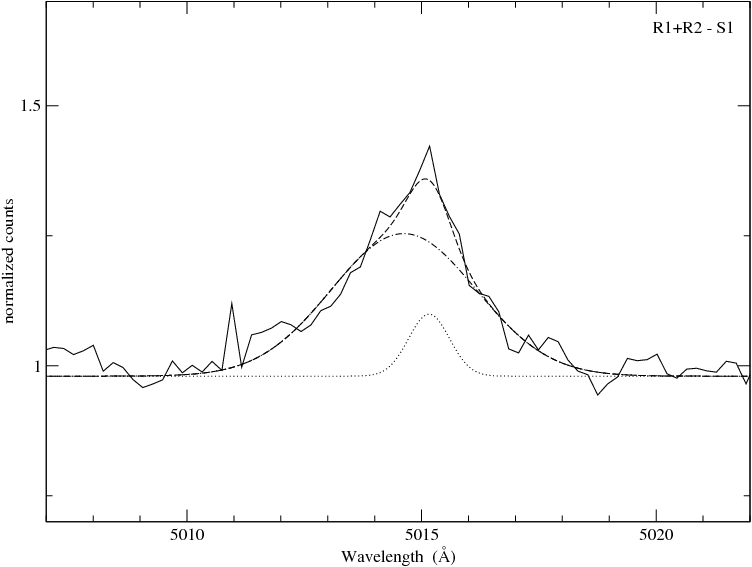}
\caption{Sections of the normalized spectrum of R1+R2 (solid line). The left
 panel shows from 4861 to 4876\,\AA\ 
 containing  H$\beta$ and the right panel shows from 5007 to 5022\,\AA\
 containing the [O{\sc iii}]\,$\lambda$\,5007\,\AA\ emission line. For both
 we have superposed the fits from the ngaussfit task in IRAF; the
 dashed-dotted line is the broad component, the dotted line is the narrow
 component and the dashed line is the sum of both.}
\label{ngauss}
\end{figure*}

In all the studied regions, however, the best fit for the
H$\beta$ line is obtained with two different components having different 
radial velocities of up to 35 km s$ ^{-1} $.  We  used the
widths of those components as an initial approximation to fit the [O{\sc
    iii}] lines which, due to their intrinsic weakness, show  lower S/N ratio,
and found them to provide also an optimal two component fit. The radial
velocities found for 
the narrow (broad) components of both H$\beta$ and [O{\sc iii}] are the
same within the errors. An example of the two-Gaussian fit is shown in Fig.
\ref{ngauss}. 

For each CNSFR, the gas velocity dispersions for the H$\beta$ and [O{\sc
iii}]\,$\lambda$\,5007\,\AA\ lines derived using single and double line
Gaussian fit, and their corresponding errors are listed in Table
\ref{disp}. Columns 4 and 5, labelled `One component', give the results for
the single Gaussian fit. Columns 6 and 7, and 8 and 9,  labelled `Two
components - Narrow' and `Two components - Broad', respectively,  list the
results for the two component fits. The last column of the table, labelled
$\Delta$v$_{nb}$, gives the velocity difference between the narrow and broad
components. This is calculated as the average of the H$\beta$ and [O{\sc iii}]
fit differences. Taking into account the errors in the two component fits, the
errors in these velocity differences vary from  10 to 15\,km\,s$^{-1}$.

We have also determined the distribution along each slit position of the
radial velocities and the velocity dispersions of the ionized gas using the
same procedure as in the case of the stars, that is using spectra extracted
every two pixels for S1 and every three pixels, superposing one pixel for
consecutive extractions, for S2. These spectra, however, do not have the
required S/N ratio to allow an acceptable two-component fit, therefore a
single-Gaussian component has been used. The goodness of this procedure is
discussed below in Section 4.

\subsection{Emission line ratios}
\label{lineratios}

\begin{figure*}
\includegraphics[width=.48\textwidth,height=.30\textwidth,angle=0]{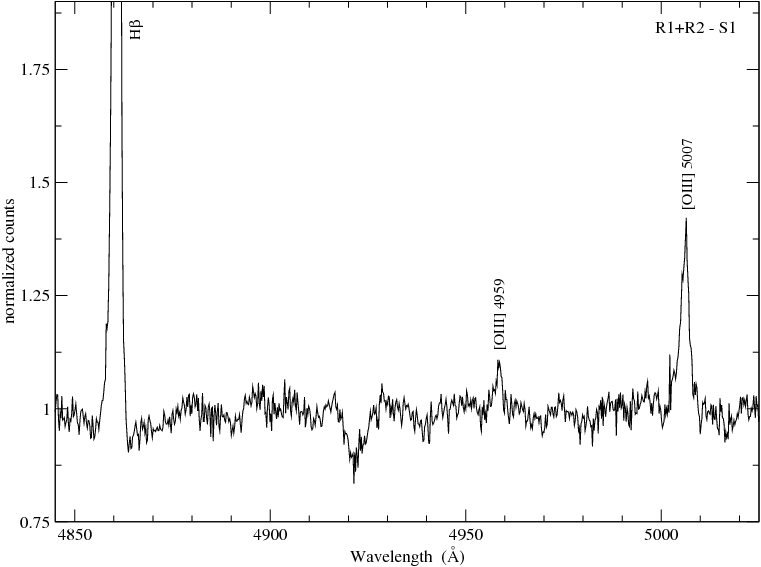}
\includegraphics[width=.48\textwidth,height=.30\textwidth,angle=0]{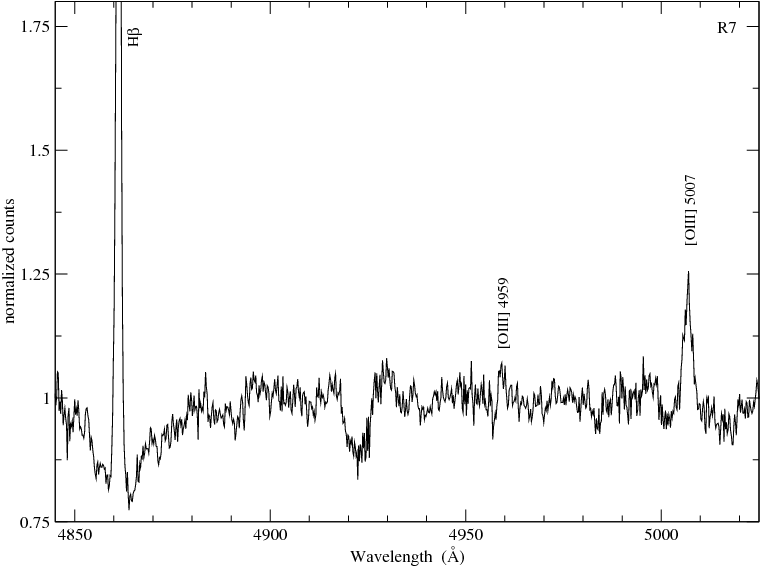}
\caption{Enlargement of the blue rest frame normalized spectra of R1+R2 (left)
  and R7 (right).}
\label{enlarg}
\end{figure*}

We have used two different ways to integrate the intensity of a given line:
(1) if an adequate fit was attained by a single Gaussian, the emission line 
intensities were measured using the SPLOT task in IRAF. For the H$\beta$
emission lines a conspicuous underlying stellar population is inferred
from the presence of absorption features that depress the lines
\cite[e.g.~see discussion in][]{1988MNRAS.231...57D}. Examples of this effect 
can
be appreciated in Fig.\ \ref{enlarg}. We have defined a pseudo-continuum
at the base  of the line to measure the line intensities and minimize
the errors introduced by the underlying population \cite[for details
  see][]{2006MNRAS.372..293H}. (2) When the optimal fit was obtained by two
Gaussians the individual intensities  of the narrow and broad components
are estimated from the fitting parameters
(I\,=\,1.0645\,A\,$\times$\,FWHM\,=\,$\sqrt{2\pi}$\,A\,$\sigma$; where I is
the Gaussian intensity, A is the amplitude of the Gaussian, FWHM is the full
width at half-maximum and $\sigma$ is the dispersion of the Gaussian). 
A pseudo-continuum for the H$\beta$ emission line was also defined in these
cases. The statistical errors associated with the observed
emission fluxes have been calculated with the expression
$\sigma_{l}$\,=\,$\sigma_{c}$N$^{1/2}$[1 + EW/(N$\Delta$)]$^{1/2}$, where
$\sigma_{l}$ is  the error in the observed line flux, $\sigma_{c}$ represents
the standard 
deviation in a box near the measured emission line and stands for the error in
the continuum placement, N is the number of pixels used in the measurement of 
the line intensity, EW is the line equivalent width and $\Delta$ is 
the wavelength dispersion in \AA\,pixel$^{-1}$ \citep{1994ApJ...437..239G}. 
For the H$\beta$ emission line we have
doubled the derived error, $\sigma_{l}$, in order to take into account the
uncertainties introduced by the presence of the underlying stellar population 
\citep{2006MNRAS.372..293H}. 

The logarithmic ratio between the emission line intensities of [O{\sc
iii}]\,$\lambda$\,5007\,\AA\ and H$\beta$ and their corresponding errors
are presented in Table \ref{ratios}. We have also listed the
logarithmic ratio between the emission line fluxes of [N{\sc
ii}]\,$\lambda$\,6584\,\AA\ and H$\alpha$ together with their corresponding
errors from \cite{2007MNRAS.382..251D} for R1+R2 and R4, and from
\cite{1997A&A...325...81P} for R7 and the nucleus.
For these last two objects, the [N{\sc ii}] and H$\alpha$ emission are
derived from the H$\alpha$+[N{\sc ii}] narrow band images. Planesas and
collaborators estimated the relative contribution of the [N{\sc
    ii}]\,$\lambda\lambda$\,6548,6584\,\AA\ lines from the H$\alpha$ and
[N{\sc ii}]\,$\lambda$\,6584\,\AA\ equivalent width measurements of
\cite{1982ApJS...50..517S}. The logarithmic ratios for R1+R2
and R4 derived using this procedure are similar
(-0.64\,$\pm$\,0.04 and -0.65\,$\pm$\,0.04). For the X region we assumed a
value of -0.38 (without error) for this ratio as given
by D\'iaz et al.\ (2007) for the other CNSFRs of this galaxy.


\begin{table*}
\centering
\caption[]{Line ratios.}
\begin{tabular} {l c c c c c}
\hline
        &      &         One component &
        \multicolumn{2}{c}{Two components}   & \\
        &      &    &  Narrow  & Broad &  \\
 Region & Slit & log([O{\sc iii}]5007/H$\beta$) & log([O{\sc
        iii}]5007/H$\beta$) & log([O{\sc iii}]5007/H$\beta$) & log([N{\sc
        ii}]6584/H$\alpha$) \\
\hline

R1+R2 &  S1   &  -1.03$\pm$0.05   & -1.64$\pm$0.09  & -0.76$\pm$0.10 & -0.37$\pm$0.01$^a$\\
R1+R2 &  S2   &  -0.92$\pm$0.07   & -1.51$\pm$0.11  & -0.57$\pm$0.14 &                   \\
R4    &  S2   &  -0.78$\pm$0.11   & -1.01$\pm$0.10  & -0.71$\pm$0.15 & -0.38$\pm$0.01$^a$\\
R7    &  S1   &  -0.78$\pm$0.10   & -1.79$\pm$0.12  & -0.40$\pm$0.18 & -0.66$\pm$0.04$^b$\\[2pt]
N     &  S2   &  -0.50$\pm$0.07   & -0.92$\pm$0.10  & -0.23$\pm$0.18 & -0.68$\pm$0.04$^b$\\[2pt]
X     &  S1   &  -0.93$\pm$0.06   & -1.45$\pm$0.09  & -0.77$\pm$0.11 & -0.38:$^c$ \\

\hline

\multicolumn{6}{@{}l}{$^a$from \citet{2007MNRAS.382..251D}}\\
\multicolumn{6}{@{}l}{$^b$from Planesas et al.\ (1997)}\\
\multicolumn{6}{@{}l}{$^c$Assumed from the values given by D\'iaz et al.\ (2007) for the other observed CNSFR in NGC\,2903}
\end{tabular}
\label{ratios}
\end{table*}



\begin{figure*}
\centering
\begin{minipage}[c]{6.1in}
\includegraphics[width=.68\textwidth,angle=90]{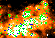}
\includegraphics[width=.68\textwidth,angle=90]{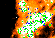}\\
\includegraphics[width=.68\textwidth,angle=90]{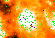}
\includegraphics[width=.68\textwidth,angle=90]{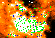}\\
\end{minipage}
\begin{minipage}[c]{0.5in}
\rotcaption{Enlargements of the F606W image around the CNSFRs of our study with
  the contours overlapped. The circles correspond to the adopted radius
  for each region. 
[{\it See the electronic edition of the Journal for a  
  colour version of this figure where the adopted radii are in blue and
  the contours corresponding to the half light brightness are in red.}]} 
\label{sizes}
\end{minipage}
\end{figure*}

\section{Dynamical mass derivation}
The mass of a virialized stellar system is given by two parameters: its
velocity dispersion and its size.

In order to determine the sizes of the stellar clusters within our observed
CNSFRs, we have used the retrieved wide V HST image which provides a spatial
resolution of 0.045 \arcsec per pixel. At the adopted distance for NGC\,2903, this
resolution corresponds to 1.9\,pc/pixel. Fig.\ \ref{sizes} shows enlargements
around the different regions studied with intensity contours overlapped.

We find, as expected, that the four CNSFRs studied here are formed by a large
number of individual star-forming clusters. Regions
R1 and R2 form an interrelated complex with two main structures: R1 and R2
surrounded by, at least, 31 secondary structures. R1 is resolved into two
separate clusters, labelled R1a and R1b and shows other two associated knots,
labelled R1c and R1d. R2 is made up of a main cluster, labelled R2, and four
secondary ones labelled ``a'' to ``d''. The rest of the knots can not be
directly 
related to any of the two main structures, and have been labelled as R12
followed by a letter from ``a'' to ``v''. The same criterion has been applied
for 
regions R4 and R7 for which we find 28 and 27 individual clusters,
respectively. On the other hand, the galaxy nucleus seems to be formed by one
main structure with two small secondary ones to the South West, and some very
weak knots which are difficult to appreciate in the picture. The secondary
structures have been designated by Na and Nb. In all cases the knots have been
found with a detection 
level of 10\,$\sigma$ above the background. All these knots are within the
radius of the regions defined by Planesas et al.\ (1997). We have to remark
that our search for knots has not been exhaustive since that 
is not the aim of this work.



\begin{table*}
\centering
\caption[]{Positions, radii and peak intensities derived from the HST-F606W image.}
\begin{tabular}{@{}lcccc@{}p{0.6cm}@{}lcccc@{}}
\cline{1-5}\cline{7-11} 
Region &  \multicolumn{2}{c}{position} &  R & I & & Region &  \multicolumn{2}{c}{position} &  R & I \\
  &  $\alpha_{J2000.0}$ & $\delta_{J2000.0}$ & (pc) & (counts) & & &  $\alpha_{J2000.0}$ & $\delta_{J2000.0}$ & (pc) & (counts) \\
\cline{1-5}\cline{7-11}

R1a       & 9$^{\romn h}$32$^{\romn m}$10\fs06 & +21$^\circ$30\arcmin06\farcs86 & 1.6$\pm$0.1 & 4083 & & R4n       & 9$^{\romn h}$32$^{\romn m}$10\fs09 & +21$^\circ$29\arcmin59\farcs86 & 1.5$\pm$0.3 &  240\\
R1b       & 9$^{\romn h}$32$^{\romn m}$10\fs06 & +21$^\circ$30\arcmin06\farcs95 & 1.6$\pm$0.2 & 4015 & & R4o       & 9$^{\romn h}$32$^{\romn m}$10\fs18 & +21$^\circ$29\arcmin59\farcs70 & 2.3$\pm$0.3 &  237\\
R1c       & 9$^{\romn h}$32$^{\romn m}$10\fs07 & +21$^\circ$30\arcmin07\farcs10 & 1.5$\pm$0.1 &  363 & & R4p       & 9$^{\romn h}$32$^{\romn m}$10\fs16 & +21$^\circ$29\arcmin59\farcs10 & 1.6$\pm$0.2 &  236\\
R1d       & 9$^{\romn h}$32$^{\romn m}$10\fs05 & +21$^\circ$30\arcmin07\farcs08 & 2.1$\pm$0.1 &  254 & & R4q       & 9$^{\romn h}$32$^{\romn m}$10\fs25 & +21$^\circ$29\arcmin58\farcs92 & 2.7$\pm$0.3 &  228\\
R2        & 9$^{\romn h}$32$^{\romn m}$09\fs98 & +21$^\circ$30\arcmin06\farcs65 & 1.9$\pm$0.1 & 2892 & & R4r       & 9$^{\romn h}$32$^{\romn m}$10\fs10 & +21$^\circ$29\arcmin59\farcs71 & 1.7$\pm$0.2 &  227\\
R2a       & 9$^{\romn h}$32$^{\romn m}$09\fs96 & +21$^\circ$30\arcmin06\farcs28 & 1.9$\pm$0.1 &  390 & & R4s       & 9$^{\romn h}$32$^{\romn m}$10\fs21 & +21$^\circ$30\arcmin00\farcs18 & 2.7$\pm$0.3 &  216\\
R2b       & 9$^{\romn h}$32$^{\romn m}$09\fs96 & +21$^\circ$30\arcmin06\farcs61 & 1.9$\pm$0.2 &  307 & & R4t       & 9$^{\romn h}$32$^{\romn m}$10\fs08 & +21$^\circ$29\arcmin59\farcs83 & 1.8$\pm$0.2 &  208\\
R2c       & 9$^{\romn h}$32$^{\romn m}$09\fs96 & +21$^\circ$30\arcmin07\farcs00 & 1.9$\pm$0.1 &  209 & & R4u       & 9$^{\romn h}$32$^{\romn m}$10\fs11 & +21$^\circ$29\arcmin58\farcs94 & 2.4$\pm$0.3 &  198\\
R2d       & 9$^{\romn h}$32$^{\romn m}$09\fs98 & +21$^\circ$30\arcmin06\farcs30 & 1.8$\pm$0.2 &  206 & & R4v       & 9$^{\romn h}$32$^{\romn m}$10\fs20 & +21$^\circ$30\arcmin00\farcs37 & 2.1$\pm$0.3 &  196\\
R12a      & 9$^{\romn h}$32$^{\romn m}$09\fs96 & +21$^\circ$30\arcmin07\farcs68 & 1.9$\pm$0.1 &  995 & & R4w       & 9$^{\romn h}$32$^{\romn m}$10\fs23 & +21$^\circ$30\arcmin00\farcs67 & 2.0$\pm$0.2 &  195\\
R12b      & 9$^{\romn h}$32$^{\romn m}$10\fs09 & +21$^\circ$30\arcmin05\farcs94 & 2.5$\pm$0.1 &  580 & & R4x       & 9$^{\romn h}$32$^{\romn m}$10\fs22 & +21$^\circ$30\arcmin01\farcs17 & 3.2$\pm$0.4 &  194\\
R12c      & 9$^{\romn h}$32$^{\romn m}$10\fs08 & +21$^\circ$30\arcmin06\farcs03 & 2.5$\pm$0.1 &  574 & & R4y       & 9$^{\romn h}$32$^{\romn m}$10\fs12 & +21$^\circ$29\arcmin58\farcs82 & 1.9$\pm$0.3 &  187\\
R12d      & 9$^{\romn h}$32$^{\romn m}$10\fs08 & +21$^\circ$30\arcmin05\farcs42 & 2.1$\pm$0.1 &  427 & & R4z       & 9$^{\romn h}$32$^{\romn m}$10\fs19 & +21$^\circ$30\arcmin00\farcs25 & 2.0$\pm$0.3 &  180\\
R12e      & 9$^{\romn h}$32$^{\romn m}$10\fs15 & +21$^\circ$30\arcmin06\farcs80 & 1.6$\pm$0.1 &  381 & & R4$\alpha$& 9$^{\romn h}$32$^{\romn m}$10\fs24 & +21$^\circ$30\arcmin00\farcs43 & 2.0$\pm$0.3 &  173\\
R12f      & 9$^{\romn h}$32$^{\romn m}$10\fs03 & +21$^\circ$30\arcmin06\farcs38 & 1.8$\pm$0.1 &  341 & & R7        & 9$^{\romn h}$32$^{\romn m}$10\fs38 & +21$^\circ$30\arcmin10\farcs93 & 2.7$\pm$0.1 &  984\\
R12g      & 9$^{\romn h}$32$^{\romn m}$10\fs04 & +21$^\circ$30\arcmin06\farcs51 & 3.2$\pm$0.2 &  293 & & R7a       & 9$^{\romn h}$32$^{\romn m}$10\fs32 & +21$^\circ$30\arcmin09\farcs74 & 2.7$\pm$0.2 &  715\\
R12h      & 9$^{\romn h}$32$^{\romn m}$10\fs04 & +21$^\circ$30\arcmin05\farcs91 & 3.6$\pm$0.4 &  266 & & R7b       & 9$^{\romn h}$32$^{\romn m}$10\fs27 & +21$^\circ$30\arcmin09\farcs99 & 3.4$\pm$0.1 &  579\\
R12i      & 9$^{\romn h}$32$^{\romn m}$10\fs15 & +21$^\circ$30\arcmin05\farcs45 & 3.2$\pm$0.3 &  255 & & R7c       & 9$^{\romn h}$32$^{\romn m}$10\fs34 & +21$^\circ$30\arcmin09\farcs65 & 2.7$\pm$0.3 &  520\\
R12j      & 9$^{\romn h}$32$^{\romn m}$10\fs13 & +21$^\circ$30\arcmin05\farcs42 & 3.2$\pm$0.4 &  238 & & R7d       & 9$^{\romn h}$32$^{\romn m}$10\fs39 & +21$^\circ$30\arcmin09\farcs17 & 2.6$\pm$0.2 &  470\\
R12k      & 9$^{\romn h}$32$^{\romn m}$10\fs11 & +21$^\circ$30\arcmin05\farcs54 & 4.0$\pm$0.5 &  233 & & R7e       & 9$^{\romn h}$32$^{\romn m}$10\fs26 & +21$^\circ$30\arcmin09\farcs56 & 1.5$\pm$0.2 &  283\\
R12l      & 9$^{\romn h}$32$^{\romn m}$10\fs03 & +21$^\circ$30\arcmin05\farcs40 & 1.9$\pm$0.1 &  219 & & R7f       & 9$^{\romn h}$32$^{\romn m}$10\fs35 & +21$^\circ$30\arcmin09\farcs03 & 4.0$\pm$0.5 &  270\\
R12m      & 9$^{\romn h}$32$^{\romn m}$10\fs06 & +21$^\circ$30\arcmin06\farcs24 & 3.8$\pm$0.3 &  212 & & R7g       & 9$^{\romn h}$32$^{\romn m}$10\fs29 & +21$^\circ$30\arcmin09\farcs64 & 2.5$\pm$0.2 &  259\\
R12n      & 9$^{\romn h}$32$^{\romn m}$10\fs01 & +21$^\circ$30\arcmin07\farcs35 & 1.8$\pm$0.1 &  209 & & R7h       & 9$^{\romn h}$32$^{\romn m}$10\fs29 & +21$^\circ$30\arcmin09\farcs76 & 2.2$\pm$0.2 &  252\\
R12o      & 9$^{\romn h}$32$^{\romn m}$10\fs01 & +21$^\circ$30\arcmin06\farcs36 & 2.6$\pm$0.2 &  207 & & R7i       & 9$^{\romn h}$32$^{\romn m}$10\fs37 & +21$^\circ$30\arcmin10\farcs54 & 2.3$\pm$0.1 &  248\\
R12p      & 9$^{\romn h}$32$^{\romn m}$10\fs01 & +21$^\circ$30\arcmin06\farcs23 & 2.8$\pm$0.2 &  204 & & R7j       & 9$^{\romn h}$32$^{\romn m}$10\fs31 & +21$^\circ$30\arcmin10\farcs24 & 2.6$\pm$0.2 &  246\\
R12q      & 9$^{\romn h}$32$^{\romn m}$10\fs12 & +21$^\circ$30\arcmin05\farcs23 & 2.3$\pm$0.2 &  203 & & R7k       & 9$^{\romn h}$32$^{\romn m}$10\fs30 & +21$^\circ$30\arcmin09\farcs80 & 2.6$\pm$0.3 &  234\\
R12r      & 9$^{\romn h}$32$^{\romn m}$10\fs11 & +21$^\circ$30\arcmin05\farcs89 & 2.7$\pm$0.4 &  198 & & R7l       & 9$^{\romn h}$32$^{\romn m}$10\fs34 & +21$^\circ$30\arcmin09\farcs21 & 3.2$\pm$0.5 &  231\\
R12s      & 9$^{\romn h}$32$^{\romn m}$10\fs05 & +21$^\circ$30\arcmin06\farcs17 & 2.1$\pm$0.1 &  193 & & R7m       & 9$^{\romn h}$32$^{\romn m}$10\fs26 & +21$^\circ$30\arcmin10\farcs19 & 2.8$\pm$0.2 &  230\\
R12t      & 9$^{\romn h}$32$^{\romn m}$10\fs16 & +21$^\circ$30\arcmin05\farcs67 & 3.5$\pm$0.4 &  187 & & R7n       & 9$^{\romn h}$32$^{\romn m}$10\fs35 & +21$^\circ$30\arcmin10\farcs06 & 2.1$\pm$0.3 &  227\\
R12u      & 9$^{\romn h}$32$^{\romn m}$10\fs15 & +21$^\circ$30\arcmin05\farcs81 & 2.1$\pm$0.2 &  180 & & R7o       & 9$^{\romn h}$32$^{\romn m}$10\fs33 & +21$^\circ$30\arcmin09\farcs47 & 2.1$\pm$0.2 &  226\\
R12v      & 9$^{\romn h}$32$^{\romn m}$10\fs14 & +21$^\circ$30\arcmin06\farcs07 & 1.7$\pm$0.2 &  179 & & R7p       & 9$^{\romn h}$32$^{\romn m}$10\fs33 & +21$^\circ$30\arcmin10\farcs25 & 1.9$\pm$0.1 &  225\\
R4        & 9$^{\romn h}$32$^{\romn m}$10\fs21 & +21$^\circ$29\arcmin59\farcs66 & 3.4$\pm$0.1 & 3772 & & R7q       & 9$^{\romn h}$32$^{\romn m}$10\fs32 & +21$^\circ$30\arcmin10\farcs56 & 1.7$\pm$0.1 &  222\\
R4a       & 9$^{\romn h}$32$^{\romn m}$10\fs08 & +21$^\circ$30\arcmin00\farcs41 & 1.9$\pm$0.2 &  727 & & R7r       & 9$^{\romn h}$32$^{\romn m}$10\fs32 & +21$^\circ$30\arcmin09\farcs35 & 1.9$\pm$0.2 &  217\\
R4b       & 9$^{\romn h}$32$^{\romn m}$10\fs15 & +21$^\circ$29\arcmin59\farcs23 & 1.7$\pm$0.2 &  711 & & R7s       & 9$^{\romn h}$32$^{\romn m}$10\fs36 & +21$^\circ$30\arcmin09\farcs27 & 3.8$\pm$0.5 &  215\\
R4c       & 9$^{\romn h}$32$^{\romn m}$10\fs12 & +21$^\circ$29\arcmin59\farcs45 & 1.9$\pm$0.1 &  703 & & R7t       & 9$^{\romn h}$32$^{\romn m}$10\fs34 & +21$^\circ$30\arcmin10\farcs35 & 2.1$\pm$0.4 &  212\\
R4d       & 9$^{\romn h}$32$^{\romn m}$10\fs11 & +21$^\circ$29\arcmin59\farcs43 & 2.7$\pm$0.2 &  673 & & R7u       & 9$^{\romn h}$32$^{\romn m}$10\fs38 & +21$^\circ$30\arcmin10\farcs57 & 2.9$\pm$0.4 &  209\\
R4e       & 9$^{\romn h}$32$^{\romn m}$10\fs21 & +21$^\circ$30\arcmin00\farcs93 & 3.4$\pm$0.4 &  476 & & R7v       & 9$^{\romn h}$32$^{\romn m}$10\fs39 & +21$^\circ$30\arcmin10\farcs60 & 2.2$\pm$0.2 &  208\\
R4f       & 9$^{\romn h}$32$^{\romn m}$10\fs14 & +21$^\circ$29\arcmin59\farcs88 & 2.9$\pm$0.1 &  429 & & R7w       & 9$^{\romn h}$32$^{\romn m}$10\fs34 & +21$^\circ$30\arcmin08\farcs86 & 1.9$\pm$0.2 &  208\\
R4g       & 9$^{\romn h}$32$^{\romn m}$10\fs17 & +21$^\circ$30\arcmin00\farcs05 & 2.9$\pm$0.4 &  310 & & R7x       & 9$^{\romn h}$32$^{\romn m}$10\fs32 & +21$^\circ$30\arcmin10\farcs04 & 2.4$\pm$0.2 &  207\\
R4h       & 9$^{\romn h}$32$^{\romn m}$10\fs16 & +21$^\circ$30\arcmin00\farcs35 & 1.5$\pm$0.2 &  279 & & R7y       & 9$^{\romn h}$32$^{\romn m}$10\fs34 & +21$^\circ$30\arcmin08\farcs95 & 1.7$\pm$0.1 &  206\\
R4i       & 9$^{\romn h}$32$^{\romn m}$10\fs21 & +21$^\circ$30\arcmin00\farcs09 & 2.0$\pm$0.3 &  274 & & R7z       & 9$^{\romn h}$32$^{\romn m}$10\fs28 & +21$^\circ$30\arcmin10\farcs30 & 2.2$\pm$0.1 &  204\\
R4j       & 9$^{\romn h}$32$^{\romn m}$10\fs18 & +21$^\circ$29\arcmin59\farcs54 & 2.4$\pm$0.2 &  271 & & N         & 9$^{\romn h}$32$^{\romn m}$10\fs14 & +21$^\circ$30\arcmin03\farcs04 & 3.8$\pm$0.1 &  870\\
R4k       & 9$^{\romn h}$32$^{\romn m}$10\fs27 & +21$^\circ$29\arcmin58\farcs99 & 2.0$\pm$0.2 &  253 & & Na        & 9$^{\romn h}$32$^{\romn m}$10\fs11 & +21$^\circ$30\arcmin02\farcs88 & 1.5$\pm$0.2 &  186\\
R4l       & 9$^{\romn h}$32$^{\romn m}$10\fs10 & +21$^\circ$29\arcmin59\farcs78 & 2.1$\pm$0.3 &  246 & & Nb        & 9$^{\romn h}$32$^{\romn m}$10\fs12 & +21$^\circ$30\arcmin02\farcs75 & 2.1$\pm$0.2 &  181\\
R4m       & 9$^{\romn h}$32$^{\romn m}$10\fs14 & +21$^\circ$29\arcmin59\farcs52 & 2.2$\pm$0.3 &  245 & & \\
\cline{1-5}\cline{7-11} 
\end{tabular}
\label{knots}
\end{table*}


We have fitted circular regions to the intensity contours corresponding to the
half light brightness distribution of each single structure (see
Fig.\ \ref{sizes}), following the procedure given in
\cite{1995AJ....110.2665M}, assuming that the regions have a circularly
symmetric Gaussian profile. The radii of the single knots vary between 1.5 and
4.0\,pc. Table \ref{knots} gives, for each identified knot, the position, as
given by the astrometric calibration of the HST image; the radius of the
circular region defined as described above together with its error;  and the
peak intensity in counts, as measured from the WFPC2 image. The nucleus of
NGC\,2903 is rather compact, but resolved, with a radius of
3.8\,pc. Obviously, the spatial resolution and, at 
least, the estimates of the smaller radii depend on the distance to each
galaxy.



\label{masses}


Upper limits to the dynamical masses (M$_{\ast}$) inside the half-light
radius (R) for each observed knot have been estimated under the
following assumptions: (i) the systems are spherically symmetric; (ii) they 
are gravitationally bound; and (iii) they have isotropic velocity
distributions [$\sigma^2$(total)\,=\,3 $\sigma_{\ast}^2$]. 
{The general expression for the virial mass of a cluster is
$\eta$\,$\sigma_{\ast}^2$\,R/G, where R is the effective gravitational
radius and $\eta$ is a dimensionless number that takes into account departures 
from isotropy in the velocity distribution and the spatial mass distribution, binary
fraction, mean surface density,
etc.\ \citep{2005ApJ...620L..27B,2006MNRAS.369.1392F}. Following
\cite{1996ApJ...466L..83H,1996ApJ...472..600H}, and for consistence with Paper~I
and \cite{tesisguiye}, we obtain the dynamical masses inside the half-light radius
using $\eta$\,=\,3 and adopting the half-light radius as a reasonable
approximation of the effective radius. Other authors
\citep[e.g.][]{1987degc.book.....S,2001MNRAS.326.1027S,2008A&A...490..125M}
assumed that the $\eta$ value is about 9.75 thus obtaining the total mass. 
On the absence of any knowledge about the tidal radius of the clusters, we
adopted this conservative approach. On the derived
masses, the different adopted $\eta$ values act as multiplicative factors.
}

It must be noted that we have measurements for the size of each knot, but we
do not have direct access  
to the stellar velocity dispersion of each individual cluster, since
our spectroscopic measurements encompass a wider area
(1.0\,$\times$\,1.9\,arcsec$^2$, which corresponds approximately to
42\,$\times$\,76\,pc$^2$ at the adopted distance of NGC\,2903) that includes
the whole  CNSFRs to which each group of knots belong.

The estimated dynamical masses for each knot and their corresponding errors  
are listed in Table \ref{mass}. For the regions that have been observed in
more than one slit position, we list the derived values using the two
separate stellar velocity dispersions. The dynamical masses in the rows
labelled  ``sum'' have been found by adding the individual masses in a given
CNSFR: R12, R4 and R7, as well as the galaxy nucleus, N.  We have taken the
average of the R12sum 
estimated from slit positions S1 and S2 as the ``adopted'' dynamical mass for R12.
The fractional errors of the dynamical masses of the individual knots and of 
the CNSFRs are listed in column 4. 


\begin{table*}
\centering
\caption[]{Dynamical masses.}
\begin{tabular} {@{}lc c c @{}p{0.5cm}@{} lc c c@{}p{0.5cm}@{} lc c c @{}}
\cline{1-4}\cline{6-9}\cline{11-14}
 Region & Slit & M$_{\ast}$ & error(\%)  & & Region & Slit & M$_{\ast}$ & error(\%)   & & Region & Slit & M$_{\ast}$ & error(\%)\\ 
  
\cline{1-4}\cline{6-9}\cline{11-14}

R1a	& S1 & 40$\pm$5   & 12 &  & R12b    & S2 & 71$\pm$6   & 9  &  & R4s     & S2 & 36$\pm$6   & 17 \\
R1b 	& S1 & 39$\pm$6   & 17 &  & R12c    & S2 & 71$\pm$6   & 9  &  & R4t     & S2 & 24$\pm$4   & 17 \\
R1c	& S1 & 37$\pm$5   & 13 &  & R12d    & S2 & 59$\pm$6   & 9  &  & R4u     & S2 & 33$\pm$6   & 18 \\
R1d	& S1 & 52$\pm$6   & 12 &  & R12e    & S2 & 45$\pm$5   & 10 &  & R4v     & S2 & 29$\pm$5   & 19 \\
R2	& S1 & 46$\pm$6   & 12 &  & R12f    & S2 & 51$\pm$5   & 10 &  & R4w     & S2 & 27$\pm$4   & 16 \\
R2a	& S1 & 47$\pm$6   & 12 &  & R12g    & S2 & 89$\pm$9   & 10 &  & R4x     & S2 & 43$\pm$8   & 18 \\
R2b	& S1 & 47$\pm$7   & 15 &  & R12h    & S2 & 101$\pm$13 & 13 &  & R4y     & S2 & 26$\pm$5   & 20 \\
R2c	& S1 & 46$\pm$6   & 12 &  & R12i    & S2 & 89$\pm$11  & 12 &  & R4z     & S2 & 27$\pm$5   & 20 \\
R2d	& S1 & 44$\pm$7   & 16 &  & R12j    & S2 & 91$\pm$13  & 14 &  & R4$\alpha$&S2& 27$\pm$5   & 20 \\
R12a	& S1 & 47$\pm$6   & 12 &  & R12k    & S2 & 113$\pm$16 & 14 &  & R4sum   & S2 & 853$\pm$28 & 3  \\
R12b	& S1 & 63$\pm$7   & 11 &  & R12l    & S2 & 53$\pm$5   & 10 &  & R7      & S1 & 26$\pm$5   & 18 \\
R12c	& S1 & 63$\pm$7   & 11 &  & R12m    & S2 & 107$\pm$12 & 11 &  & R7a     & S1 & 26$\pm$5   & 19 \\
R12d	& S1 & 52$\pm$6   & 12 &  & R12n    & S2 & 50$\pm$5   & 10 &  & R7b     & S1 & 32$\pm$6   & 17 \\
R12e	& S1 & 40$\pm$5   & 12 &  & R12o    & S2 & 72$\pm$8   & 11 &  & R7c     & S1 & 26$\pm$5   & 20 \\
R12f	& S1 & 45$\pm$5   & 12 &  & R12p    & S2 & 79$\pm$8   & 10 &  & R7d     & S1 & 25$\pm$5   & 19 \\
R12g	& S1 & 79$\pm$10  & 12 &  & R12q    & S2 & 65$\pm$7   & 11 &  & R7e     & S1 & 14$\pm$3   & 22 \\
R12h	& S1 & 89$\pm$14  & 15 &  & R12r    & S2 & 77$\pm$12  & 15 &  & R7f     & S1 & 38$\pm$8   & 21 \\
R12i	& S1 & 79$\pm$11  & 14 &  & R12s    & S2 & 59$\pm$6   & 9  &  & R7g     & S1 & 24$\pm$5   & 19 \\
R12j	& S1 & 81$\pm$13  & 16 &  & R12t    & S2 & 100$\pm$13 & 13 &  & R7h     & S1 & 21$\pm$4   & 19 \\
R12k	& S1 & 100$\pm$16 & 16 &  & R12u    & S2 & 59$\pm$7   & 12 &  & R7i     & S1 & 22$\pm$4   & 18 \\
R12l	& S1 & 47$\pm$6   & 12 &  & R12v    & S2 & 47$\pm$6   & 13 &  & R7j     & S1 & 25$\pm$5   & 19 \\
R12m	& S1 & 94$\pm$13  & 13 &  & R12sum  & S2 & 2054$\pm$45& 2  &  & R7k     & S1 & 25$\pm$5   & 21 \\
R12n	& S1 & 44$\pm$5   & 12 &  & R12 (adopted)&&1935$\pm$66& 3  &  & R7l     & S1 & 31$\pm$7   & 23 \\
R12o	& S1 & 64$\pm$8   & 13 &  & R4      & S2 & 46$\pm$6   & 13 &  & R7m     & S1 & 27$\pm$5   & 19 \\
R12p	& S1 & 70$\pm$9   & 13 &  & R4a     & S2 & 26$\pm$4   & 16 &  & R7n     & S1 & 20$\pm$5   & 22 \\
R12q	& S1 & 58$\pm$8   & 14 &  & R4b     & S2 & 23$\pm$4   & 17 &  & R7o     & S1 & 20$\pm$4   & 20 \\
R12r	& S1 & 68$\pm$12  & 18 &  & R4c     & S2 & 26$\pm$4   & 14 &  & R7p     & S1 & 19$\pm$3   & 18 \\
R12s	& S1 & 52$\pm$6   & 12 &  & R4d     & S2 & 37$\pm$5   & 15 &  & R7q     & S1 & 17$\pm$3   & 18 \\
R12t	& S1 & 88$\pm$14  & 16 &  & R4e     & S2 & 46$\pm$8   & 17 &  & R7r     & S1 & 18$\pm$4   & 20 \\
R12u	& S1 & 52$\pm$7   & 14 &  & R4f     & S2 & 40$\pm$5   & 13 &  & R7s     & S1 & 36$\pm$8   & 22 \\
R12v	& S1 & 42$\pm$7   & 16 &  & R4g     & S2 & 40$\pm$7   & 19 &  & R7t     & S1 & 21$\pm$5   & 25 \\
R12sum	& S1 & 1816$\pm$48& 3  &  & R4h     & S2 & 20$\pm$4   & 19 &  & R7u     & S1 & 28$\pm$6   & 22 \\
R1a     & S2 & 45$\pm$5   & 10 &  & R4i     & S2 & 27$\pm$5   & 20 &  & R7v     & S1 & 21$\pm$4   & 19 \\
R1b     & S2 & 44$\pm$6   & 14 &  & R4j     & S2 & 32$\pm$5   & 15 &  & R7w     & S1 & 19$\pm$4   & 20 \\
R1c     & S2 & 41$\pm$4   & 10 &  & R4k     & S2 & 27$\pm$4   & 16 &  & R7x     & S1 & 23$\pm$4   & 19 \\
R1d     & S2 & 59$\pm$6   & 9  &  & R4l     & S2 & 29$\pm$5   & 19 &  & R7y     & S1 & 17$\pm$3   & 18 \\
R2      & S2 & 52$\pm$5   & 10 &  & R4m     & S2 & 30$\pm$6   & 18 &  & R7z     & S1 & 21$\pm$4   & 18 \\
R2a     & S2 & 53$\pm$5   & 10 &  & R4n     & S2 & 20$\pm$5   & 24 &  & R7sum   & S1 & 642$\pm$26 & 4  \\
R2b     & S2 & 53$\pm$7   & 13 &  & R4o     & S2 & 31$\pm$6   & 18 &  & N       & S2 & 112$\pm$10 & 9  \\
R2c     & S2 & 52$\pm$5   & 10 &  & R4p     & S2 & 22$\pm$4   & 18 &  & Na      & S2 & 43$\pm$7   & 16 \\
R2d     & S2 & 50$\pm$6   & 13 &  & R4q     & S2 & 36$\pm$6   & 17 &  & Nb      & S2 & 62$\pm$8   & 13 \\
R12a    & S2 & 53$\pm$5   & 10 &  & R4r     & S2 & 23$\pm$4   & 17 &  & Nsum    & S2 & 217$\pm$15 & 7  \\

\cline{1-4}\cline{6-9}\cline{11-14}
\multicolumn{9}{l}{masses in 10$^5$ M$_\odot$.}
\end{tabular}
\label{mass}
\end{table*}


\section{Ionizing star cluster properties}

For each of the CNSFR: R12, R4 and R7, the total number of ionizing photons
was derived from the total observed H$\alpha$ luminosities given by  Planesas
et al.\ (1997), 
correcting for the different assumed distance. We also corrected for internal 
extinction using the colour excess [E(B-V)] estimated by \cite{tesisdiego}
from optical spectroscopy and assuming  the  galactic 
extinction law of \cite{1972ApJ...172..593M} with $R_v$\,=\,3.2. Planesas et
al.\ (1997) estimated a diameter of 2.0\,arcsec for regions R1, R2, R7 and the
nucleus and  2.4\,arcsec for region R4. In the case of R1+R2 we added their
H$\alpha$ luminosities. No values are found in the literature for the
H$\alpha$ luminosity of region X.
Our derived values of Q(H$_0$) constitute lower limits
since we have not taken into account the fraction of photons that may have
been absorbed by dust or may have escaped the region.

Once calculated the number of Lyman continuum photons, the masses of the
ionizing star clusters, M$_{ion}$, have been derived using the solar
metallicity single burst models by \cite{1995A&AS..112...35G} which provide
the number of ionizing photons per unit mass, [$Q(H_0)/M_{ion}$]. 
A Salpeter initial mass function \citep[][IMF]{1955ApJ...121..161S} has been
assumed with lower and upper mass limits of 0.8 and 120 M$_\odot$. In order to
take into account the evolution of the \HII\ region, we have made use of the
fact that a relation exists between the degree of evolution of the cluster, as
represented by the equivalent width of the H$\beta$ emission line, and the
number of Lyman continuum photons per unit solar mass \citep[e.g.\
][]{2000MNRAS.318..462D}. We have  measured the EW(H$\beta$) from our spectra
(see Table  \ref{parameters}) following the same procedure as in
\cite{2006MNRAS.372..293H,2008MNRAS.383..209H}, that is defining a pseudo-continuum to
take into account the absorption  from the underlying stellar
population. This procedure in fact may underestimate the value of the
equivalent width, since it includes the contribution to the continuum by the
older stellar population (see discussions in D\'\i az et al. 2007 and Dors et
al.\ 2008). The derived
masses for the ionizing population of the observed CNSFRs are given in column
8 of Table \ref{parameters} and are between 1 and 4 per cent of the dynamical
mass (see column 11 of the table).


\begin{table*}
\centering
\caption[]{Physical parameters.}
\begin{tabular} {@{} l c c c c c c c c c c @{} }
\hline
 Region & L$_{obs}$(H$\alpha$)$^a$ & E(B-V)$^b$ &
 c(H$\alpha$) &L(H$\alpha$) & Q(H$_0$)  & 
 EW(H$\beta$) & M$_{ion}$ & N$_e$ & M$_{{\rm HII}}$ &
 M$_{ion}$/M$_{\ast}$   \\ 
  &  &  &  &  &  &  &  &  &  &  (per cent)  \\
\hline

R1+R2 & 20.3 & 0.53 & 0.52 &  66.3 & 48.7 & 12.1 & 18.9 & 280$^c$& 0.79 & 1.0 \\
R4    &  8.9 & 0.66 & 0.64 &  38.9 & 28.6 &  4.8 & 24.6 & 270$^c$& 0.48 & 2.9 \\
R7    &  6.4 & 0.71 & 0.69 &  31.3 & 23.0 &  3.9 & 23.6 & 350$^b$& 0.30 & 3.7 \\[2pt]
N$^d$ &  2.8 &  --- & ---  &   2.8 &  2.0 &  3.8 &  2.1 & ---    & 0.03 & 1.0 \\ 

\hline
\multicolumn{11}{l}{Note. Luminosities in 10$^{38}$\,erg\,s$^{-1}$, masses in 10$^5$
  M$_\odot$, ionizing photons in 10$^{50}$\,photon\,s$^{-1}$ and densities in
cm$^{-3}$}\\
\multicolumn{11}{l}{$^a$From Planesas et al.\ (1997) corrected for
 the different adopted distance.} \\ 
\multicolumn{11}{l}{$^b$From P\'erez-Olea (1996).}\\
\multicolumn{11}{l}{$^c$From \citet{2007MNRAS.382..251D}.}\\
\multicolumn{11}{l}{$^d$We assume a value of 0.0 and 300 for E(B-V) and N$_e$,
respectively.}
\end{tabular}
\label{parameters}
\end{table*}

The amount of ionized gas (M$_{{\rm HII}}$) associated to
each star-forming region complex can also be derived from the H$\alpha$ luminosities
using the electron density (N$_e$) dependency relation given by
\cite{1990ApJ...356..389M} for an electron temperature of 10$^4$\,K.
The electron density for each
region (obtained from  the [S{\sc ii}]\,$\lambda\lambda$\,6717\,/\,6731\,\AA\
line ratio) has been taken from \cite{2007MNRAS.382..251D} for R1+R2 and R4 and
\cite{tesisdiego} for R7 (see Table \ref{parameters}). A value of N$_e$ equal
to 300\,cm$^{-3}$, typical of nuclear \HII\ regions,  has been assumed for the nucleus.


\begin{figure}
\centering
\includegraphics[width=.44\textwidth,angle=0]{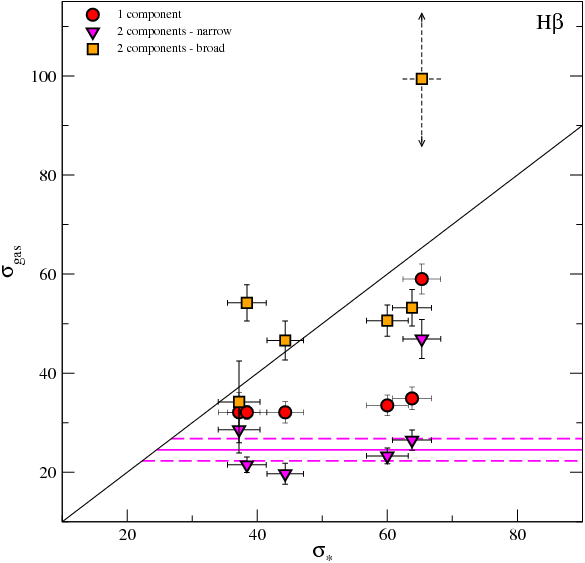}\vspace*{0.4cm}
\includegraphics[width=.44\textwidth,angle=0]{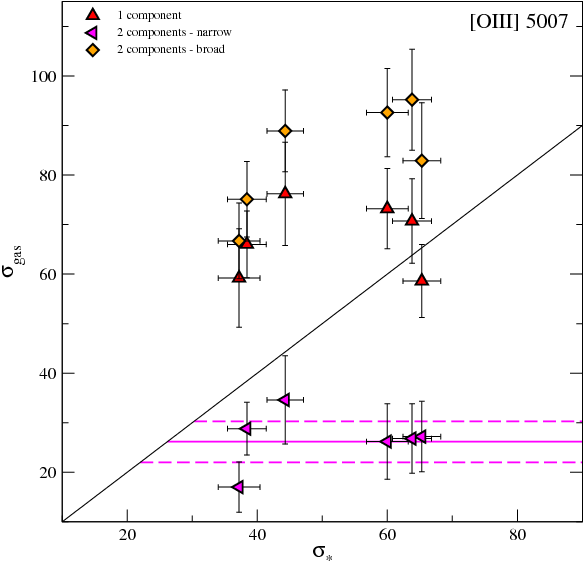}
\caption{Upper panel: relation between velocity dispersions of the gas
  (derived from H$\beta$) and stars (CaT) for the CNSFRs and the nucleus of
  NGC\,2903. Symbols are as follows: single Gaussian fit, 
(red) circles; two Gaussian fit, broad component, (orange) squares; narrow component,
  (magenta) downward triangles. Lower panel: as the upper panel for the
  [O{\sc iii}] line. (Red) upward triangles correspond to the estimates using a
  single Gaussian fit, (orange) diamonds represent the broad components of the two
  Gaussian fit and (magenta) left triangles, the narrow components. The 
  (magenta) line represents the average velocity dispersion of the narrow
  component of the gas (H$\beta$ upper and [O{\sc iii}] lower panel) for the
  CNSFRs, and the (magenta) dashed lines represent their estimated errors.[{\it
  See the electronic edition of the Journal for a   
  colour version of this figure.}]}
\label{dispersions}
\end{figure}

\section{Discussion}
\label{Velocity dispersions}
\subsection{Star and gas kinematics}

The relations between the velocity dispersions of gas, as measured from the
H$\beta$ (upper panel) and the [O{\sc iii}] (lower panel) emission lines, and
stars, as measured from the IR Ca{\sc ii} triplet, are shown in Fig.\ 
\ref{dispersions}. In both panels the straight line shows the one-to-one
relation. In the upper panel, red \footnote{In all figures, colours can be
  seen in the electronic version of the paper} 
circles show the gas velocity dispersion
measured from the H$\beta$ emission line using a single Gaussian fit. Orange
squares and magenta downward triangles show the values measured from the broad and narrow
components respectively using two-component
Gaussian fits. The deviant point, marked with arrows, corresponds to the galaxy
nucleus, whose spectra have a low signal-to-noise ratio and for which the fits do not
provide accurate results for the broad component. In the lower panel, red upward
triangles, orange diamonds and magenta left triangles correspond to the
values obtained by a single Gaussian fit, and to the broad and narrow
components of the two Gaussian fits, respectively.

In general, the H$\beta$ velocity dispersions of the CNSFRs of NGC\,2903
derived by a single Gaussian fit are lower than the stellar ones by about
25\,km\,s$^{-1}$, except for regions R7 and X, for which these values are 
very similar.  These two regions also have the lowest velocity dispersions and
in the first of them the values derived using the two different fitting
procedures are very similar. On the other hand, a much better
agreement between the velocity dispersions of stars and the broad component of
H$\beta$ for the CNSFRs is found. The narrow component shows velocity 
dispersions even lower than those obtained by single Gaussian fits, and 
similar to each other in all cases, with an average value equal to
24.5\,$\pm$\,2.2\,km\,s$^{-1}$, with the error given by the dispersion of the
individual values. This average is represented as a magenta horizontal line
in the upper panel of Fig.\ \ref{dispersions}, with its error as 
magenta dashed lines.

The [O{\sc iii}] lines however, show a behaviour different from the H$\beta$
lines. In this case, for all the CNSFRs, the gas velocity dispersions 
derived by a single Gaussian fit are higher by different amounts, between 7
and 32  \,km\,s$^{-1}$.  When a two-component fitting procedure is used, the
narrow component shows the same average value found for the H$\beta$ line,
although with a larger 
dispersion ($\sigma_g$\,=\,26.2\,$\pm$\,4.2\,km\,s$^{-1}$), while the broad
component shows values even larger than those derived from single Gaussian
fits. It is therefore apparent that, if there are two kinematically different
components in the gas, the narrow one seems to dominate the H$\beta$ line
while the broad one seems to dominate the [O{\sc iii}] ones. In fact, the
ratio between the fluxes in the narrow and broad components is between 0.65
and 0.95 for the 
H$\beta$ line (except for the weakest knot R7, for which it is about 1.46) and
decreases to between 0.06 and 0.24 for the [O{\sc iii}] line.   

These results are similar to those found for the previously analysed spiral
galaxy NGC\,3351 (see Paper I). The single Gaussian fit yields velocity
dispersions which are lower than measured for the stars in the case of
H$\beta$ and  higher in the case of [O{\sc iii}] by about 20\,km\,s$^{-1}$. 

When two widths Gaussian fits are used, the narrow components show a
relatively constant value. The average value of this velocity for the CNSFRs
analysed in both galaxies, NGC\,3351 and NGC\,2903, amounts to 23.2$\pm$
1.7\,km\,s$^{-1}$ and 25.1$\pm$2.0\,km\,s$^{-1}$ for the H$\beta$  and the
[O{\sc iii}] lines respectively. Also, for the CNSFRs in the two galaxies,
velocity dispersions associated with the broad component of H$\beta$ are in
very good agreement with the 
stellar ones, except for one deviant region in NGC\,3351 (R5). For the [O{\sc
    iii}] emission, the broad component shows velocities larger than the
stellar ones by up to about 30\,km\,s$^{-1}$.  

If the narrow component is identified with ionized
gas in a rotating disc, therefore supported by rotation, then the broad
component could, in principle, correspond to the gas response to the
gravitational potential of the stellar cluster, supported by dynamical
pressure,  explaining the coincidence with the stellar velocity dispersion in
the case of the H$\beta$ line \citep[see ][and references 
  therein]{2004A&A...424..447P}. The velocity excess shown by the broad
component of [O{\sc iii}] could be identified with peculiar velocities in the
high ionization gas related to massive star winds or even supernova remnants.

While two velocity components are clearly identified in the H$\beta$ line, the
weakness of the [O{\sc iii}] lines renders this identification more
uncertain. To test the possibility of finding a spurious result due to the low
signal-to-noise ratio in the weak [O{\sc iii}] emission lines, we  
generated a synthetic spectrum with the measured characteristics of radial
velocity and FWHM adding an artificial noise with a rms twice the observed
one. We then measured the synthetic spectra using the same technique used for
the real data, that is a double Gaussian fitting with the same initial
parameters. In all cases we obtained the same result within the observational
errors.

{ 
Our interpretation of the emission line structures, in the present study and
in Paper I, parallels that of the studies of Westmoquette and collaborators
(see for example Westmoquette et al.\ 2007a,b). They observed a narrow
($\sim$\,35-100\,km\,s$^{-1}$) and a broad ($\sim$\,100-400\,km\,s$^{-1}$)
component to the H$\alpha$ line across all their four fields in the dwarf
galaxy NGC\,1569. They conclude that the most likely explanation of the narrow
component is that it represents the general disturbed optically emitting
ionized interstellar medium (ISM), arising through a convolution of the
stirring effects of the starburst and gravitational virial motions. They also
conclude that the broad component results from the highly turbulent velocity
field associated with the interaction of the hot phase of the ISM
(material that is photo-evaporated or thermally evaporated through the action of
the strong ambient radiation field, or mechanically ablated by the impact of
fast-flowing cluster winds) with cooler gas knots, setting up turbulent mixing 
layers \citep[e.g.][]{1990MNRAS.244P..26B,1993ApJ...407...83S}.
However, our broad component velocity dispersion values derived from the
H$\beta$ emission lines resemble more their narrow component
values, since in all the CNSFRs the former are significantly lower than 
100\,km\,s$^{-1}$ and show similar values to the stellar velocity dispersions.
The [O{\sc iii}] emission line on the other hand, shows a behaviour akin
to that described by Westmoquette and collaborators.
\nocite{2007MNRAS.381..894W,2007MNRAS.381..913W}
 
There are other studies that identified an underlying broad
component to the recombination emission lines 
such as \cite{1987MNRAS.226...19D,1996MNRAS.279.1219T} in the M33 giant \HII\
region NGC\,604;
\cite*{1994ApJ...425..720C,1999MNRAS.302..677M} in the central region
of 30 Doradus; \cite{1997ApJ...488..652M} in four Wolf-Rayet galaxies, and
\cite{1999ApJ...522..199H} in the starburst galaxy NGC\,7673.
More recently, Westmoquette et al.~(2007c, 2009)
\nocite{2007ApJ...671..358W,2009arXiv0902.0064W} found a
broad feature in the H$\alpha$ emission lines in the starburst core of M82;
\cite{2007A&A...461..471O} in the blue compact galaxy ESO\,338-IG04;
\cite{2006MNRAS.370..799S,firpo2009b} in giant extragalactic \HII\ regions, and
\cite{tesisguiye,Hageleetal.2009b} in CNSFRs of early type
spiral galaxies. The first two and the last four studies also found this broad
component in the forbidden emission lines. 
}

\begin{figure*}
\includegraphics[width=.48\textwidth,height=.30\textwidth,angle=0]{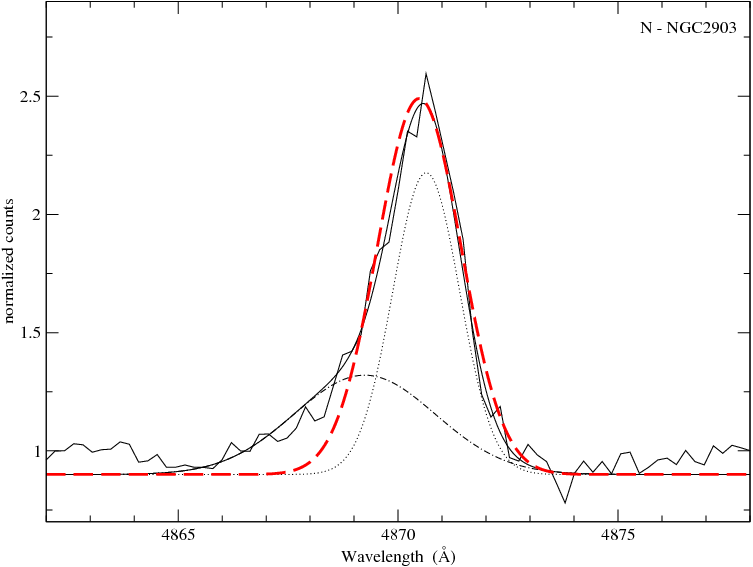}\hspace*{0.2cm}
\includegraphics[width=.48\textwidth,height=.30\textwidth,angle=0]{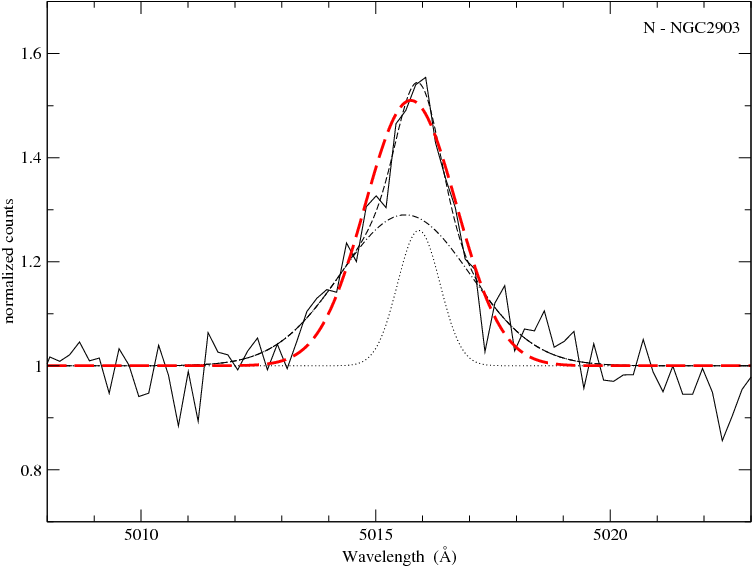}
\caption{Single and two-component Gaussian fits for the H$\beta$ (left panel)
  and [O{\sc iii}]\,$\lambda$\,5007\,\AA\ (right panel) emission lines in the
  nucleus of NGC\,2903. Thick dashed (red) line and two-component fits
  as resulting from the IRAF task ngaussfit are superposed. For the
  two-component fit, the dashed-dotted line corresponds to the broad
  component, the dotted line to the narrow component and the dashed line to
  the sum of both.} 
\label{fit-nucleus}
\end{figure*}

Single Gaussian fits yield the same
velocity dispersion in the nucleus for both emission lines, H$\beta$ and [O{\sc 
iii}]\,5007\,\AA, and close to the stellar value. If a two-component
fit is used, the low singnal-to-noise ratio in the H$\beta$ line, combined
with the underlying absorption, yields very uncertain values for the broad
component and a value for the narrow component which is closer to the stellar
one than in the case of the CNSFRs. The results for the [O{\sc
    iii}]\,5007\,\AA\ line are similar to those found for the
CNSFRs. Fig.\ \ref{fit-nucleus} shows the single and two-component Gaussian
fitting for the H$\beta$ and [O{\sc iii}]\,5007\,\AA\, lines. The $H\beta$
line shows a blue assymetry which might correspond to a
low intensity broad component. This component is barely detectable (if at all present)
in the [O{\sc iii}]\,5007\,\AA\ line. 

{
If this decomposition of the emission lines is correct, the profiles of the
nuclear lines show a behaviour similar to that described by Westmoquette and
collaborators, with the velocity dispersion value of the broad component of
H$\beta$ $\gtrsim$\,100\,km\,s$^{-1}$.  Using ROSAT data, \cite{2003A&A...411...41T} 
found a very soft X-ray emission
feature that extends to 2\arcmin\ ($\sim$\,5\,kpc) from the nucleus to the
west. They concluded that the existence of a galactic wind and its interaction
with the surrounding intergalactic medium is the most plausible source for
this soft emission.
The H$/beta$ broad feature thus, could be consistent with the presence of a nuclear region
galactic wind, which could arise from gas that is mixing into a high sound
speed, hot gas phase, whose presence dominates X-ray images, as in NGC\,1569
(Westmoquette et al.\ 2007a,b). 
}

The presence of two different gaseous components in our analysed CNSFR could
have an important effect on the classification of the activity in the central
regions of galaxies through  diagnostic diagrams and the magnitude of this
effect would increase with decreasing spatial resolution. In fact, the [O{\sc
    iii}]/H$\beta$ ratio is much smaller for the narrow component than for the
broad one, and single Gaussian fittings provide intermediate values (see Table
\ref{disp}). This seems to point to a lower excitation for the lower velocity
kinematical component. The effect is rather dramatic even for the nucleus of
NGC\,2903 for which the logarithmic [O{\sc iii}]/H$\beta$ ratio for the low
and high velocity components vary from -0.91 to -0.22. Unfortunately, we do
not have high resolution data in the [N{\sc ii}] -- H$\alpha$ wavelength range
and hence we can not investigate the complete effect in the excitation plot
(see Paper I).

\label{Radial velocities}

\begin{figure}
\centering
\includegraphics[width=.46\textwidth,angle=0]{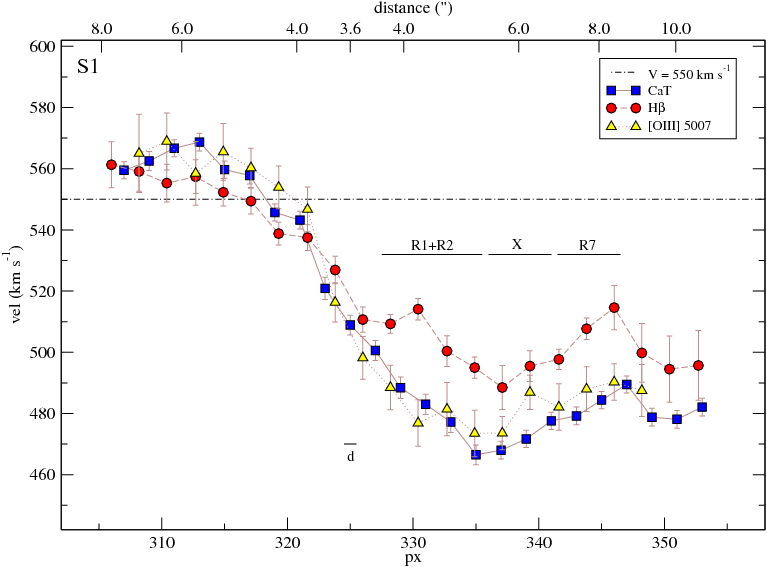}\vspace*{0.3cm}
\includegraphics[width=.46\textwidth,angle=0]{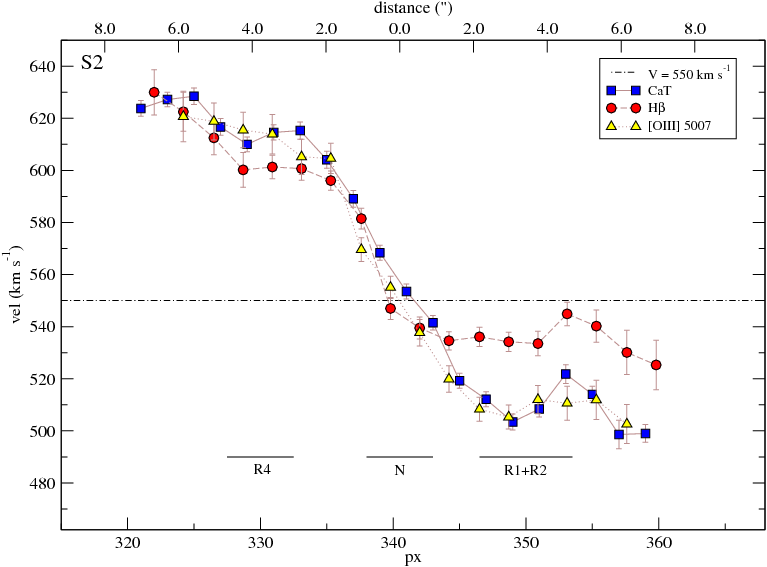}
\caption{Radial velocities along the slit versus pixel number
  for each slit position of NGC\,2903 (upper panel: S1; lower panel: S2) as
  derived from the gas emission lines (red circles: H$\beta$; upward triangles:
  [O{\sc iii}]) and the stellar absorption ones (blue squares). The individual
  CNSFRs and the nucleus, ``N'', or the closest position to it, ``d'', are
  marked in the plots. The dashed-dotted line is the systemic velocity of
  NGC\,2903 derived by Planesas et al.\ (1997). The distance in arcsec from
  the nucleus is displayed in the upper x-axis of each panel. [{\it
  See the electronic edition of the Journal for a   
  colour version of this figure.}]}
\label{velocities}
\end{figure}


The radial velocities along the slit for each angular position of NGC\,2903 as 
derived from the ionized gas emission lines, H$\beta$ and [O{\sc
iii}]\,5007\,\AA, and the stellar CaT absorptions are shown in the upper and
lower panels of Fig.\ \ref{velocities} respectively. The rotation curves seem
to have the turnover points at the same positions as the star-forming ring,
specially for the S2 slit position across the nucleus, as found in other
galaxies \citep[see][and references
  therein]{1988ApJ...334..573T,1999ApJ...512..623D}. For the systemic velocity
of NGC\,2903, the derived values are 
consistent with those previously obtained by Planesas et al.\ (1997) and
\cite{1998AJ....115...62H}, and with the velocity  distribution expected
for this type of galaxies \citep{1987gady.book.....B}.

The radial velocities derived from both the stellar CaT and the [O{\sc
    iii}]\,5007\,\AA\ emission line are in good agreement. The H$\beta$ line
velocities, however, differ from the stellar ones, more appreciably in the
Norh-West region,  in amounts  similar to the differences shown by the
radial velocities corresponding to the two Gaussian components  
($\Delta$v$_{nb}$; see Table \ref{disp}). This is compatible with the
assumption of the existence of two kinematically different components, the
broad 
one dominating the single fit in  [O{\sc iii}]  and the narrow one being the
dominant one in the case of H$\beta$.

\begin{figure}
\centering
\includegraphics[width=.46\textwidth,angle=0]{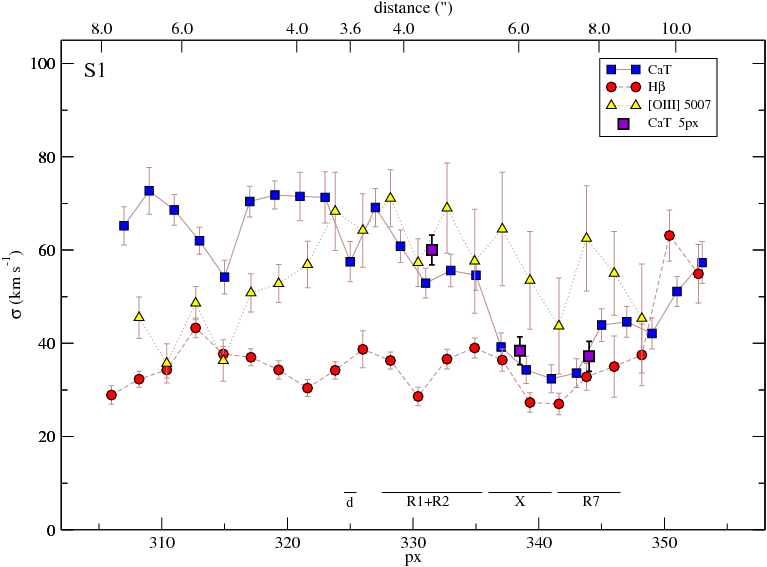}\vspace*{0.3cm}
\includegraphics[width=.46\textwidth,angle=0]{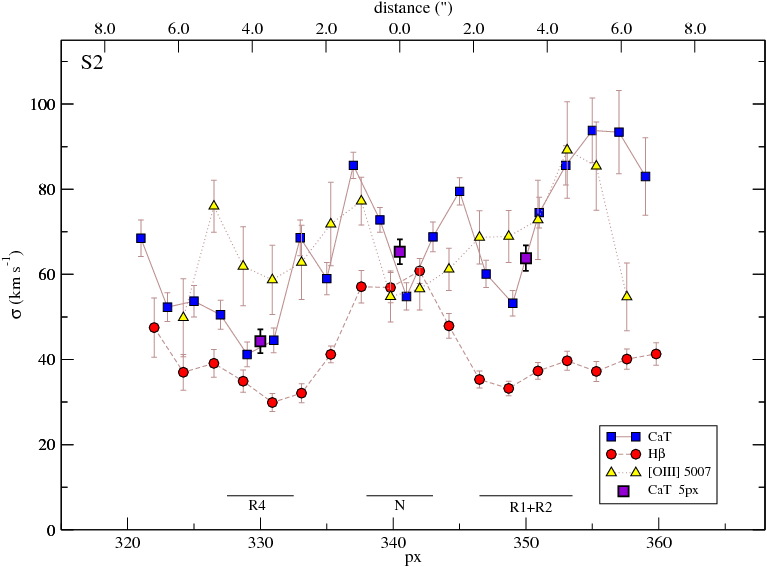}
\caption{Velocity dispersions along the slit versus pixel
  number for each slit position of NGC\,2903 (upper panel: S1; lower panel:
  S2) as derived from the gas emission lines (small red circles: H$\beta$;
  small yellow triangles: [O{\sc iii}]) and the stellar absorption ones (small
  blue squares). {The stellar velocity dispersions derived for each region
  and the nucleus using the 5\,px aperture are also plotted with violet
  squares}. The individual CNSFRs and the nucleus, ``N'', or the closest
  position to it, ``d'', are marked in the plots. The distance in arcsec from
  the nucleus is displayed in the upper x-axis of each panel. [{\it
  See the electronic edition of the Journal for a   
  colour version of this figure.}]}
\label{dispersions-px}
\end{figure}

Fig.\ \ref{dispersions-px} shows the run of velocity dispersions along the
slit versus pixel number for slit positions S1 and S2 of NGC\,2903,
respectively. These velocity dispersions have been derived from the gas
emission lines, H$\beta$ and [O{\sc iii}], and the stellar absorption ones
using 2 and 3\,px apertures for S1 and S2, respectively.
We have marked the location of the studied CNSFR and the galaxy
nucleus.
{We have also plotted the stellar velocity dispersion derived for each
region and the nucleus using the 5\,px apertures. The values
derived with wider apertures are approximately the average of the velocity
dispersions estimated using the narrower apertures. Then, the increase of the
apparent velocity dispersions of the regions lying on relatively steep parts
of the rotation curve by spatial ``beam smearing'' of the rotational velocity
gradient due to the finite angular 
resolution of the spectra is no so critical for the studied CNSFRs. This is
due to the turnover points of the rotation curves and the
star-forming ring seem to be located at the same positions, as was pointed out
above.} 
The behaviour of the velocity dispersion along slit position S1
(upper panel) is very complex. The H$\beta$ line shows a relatively constant
value of the velocity dispersion. To the SW of the R1+R2 star forming complex,
there is a clear systematic difference between the velocities measured from
the HI recombination line and the stellar absorption lines. This systematic
difference is significative lower in the region where active star formation is
taking 
place. Along slit position S2, crossing the galaxy nucleus, we detect relative
minima in the stellar and gas velocity dispersions at the positions of CNSFR
R4 and R1+R2. The stellar velocity dispersions show also a relative minimum at
the location of the nucleus although in this case it is accompanied by a rise
in the velocity dispersion measured from the H$\beta$ line.  




\subsection{Star cluster masses}

Unlike the case of the CNSFRs in NGC\,3351 studied in Paper I for which we
found that two of the observed regions, R4 and R5, seem to possess just one
knot showing up in the continuum image and coincident with the H$\alpha$
emission, the four CNSFRs observed in NGC\,2903: R1+R2, R4 and R7 show a complex
structure at the HST resolution (Fig.\ \ref{sizes}), with a good number of
subclusters with linear diameters between 3 and 8 pc. For these individual
clusters, the derived {upper limits to the} masses are in the range
between 1.4\,$\times$\,10$^6$ and  1.13\,$\times$\,10$^7$\,M$_\odot$ (see
table \ref{mass}), with fractional errors between about 9 and 25 per cent. The
{upper limits to the} dynamical masses estimated for the whole CNSFRs
(``sum'') are between 
6.4\,$\times$\,10$^7$ and  1.9\,$\times$\,10$^8$\,M$_\odot$, with fractional
errors between about 3 and 4 per cent. The {upper limit to the} dynamical
mass derived for the nuclear region inside the inner 3.8\,pc is
1.1\,$\times$\,10$^7$\,M$_\odot$, with a fractional error of about 9 per
cent.

The masses of the ionizing stellar clusters of the circumnuclear complexes,
have been 
derived from their H$\alpha$ luminosities under the assumption that the
regions are ionization bound and without taking into account any photon
absorption by dust. A range of masses  between 1.9 and
2.5\,$\times$\,10$^6$\,M$_\odot$ for the star-forming regions, and
2.1\,$\times$\,10$^5$\,M$_\odot$ for the nucleus are found (see Table
\ref{parameters}).  In column 11 of Table \ref{parameters} we show a comparison
(in percentage) between the ionizing stellar masses of the circumnuclear regions
and their dynamical masses. These values are approximately between 1\,-\,4 per
cent for the CNSFRs, and  1 per cent for the nucleus of NGC\,2903.

Finally, the masses of the ionized gas, also derived from their H$\alpha$
luminosities, range between 3\,$\times$\,10$^4$ and
8\,$\times$\,10$^5$\,M$_\odot$ for the CNSFRs, and
3\,$\times$\,10$^3$\,M$_\odot$ for the nucleus (see Table
\ref{parameters}). They make up a very small fraction of the total mass of the
regions. Both the masses of the ionizing stellar clusters and those of the
ionized gas are comparable to those 
derived by \cite{1995ApJ...439..604G} for the circumnuclear region A in
NGC\,7714. It should be taken into account that the latter
have been derived from
the H$\alpha$ luminosity of the CNSFRs assuming a single kinematical component
for the emission line. 
If we consider only the broad component whose
kinematics follows that of the stars in the regions, all derived quantities
would be smaller by a factor of 2.

The circumnuclear star formation in NGC\,2903 has been studied by
\cite{2001MNRAS.322..757A} using high resolution near-IR photometry in the H
band and ground based IR spectroscopy. Although, in general, the spatial
distribution of the stellar clusters detected in the IR does not coincide with
the maxima of the H$\alpha$ emission, some CNSFR show up prominently in both
spectral ranges. This is the case for our regions R4, R7 and R1+R2 which can
be identified with regions H8, H4 and H1+H2, respectively, in
\cite{2001MNRAS.322..757A}. These authors have estimated the masses of the
young clusters present in the central region of NGC\,2903 from the NIR colours
using stellar population synthesis models corresponding to single bursts of
star formation with Gaussian FWHM values of 1, 5 and 100\,Myr and a Salpeter
IMF with lower and upper mass cut-offs of 1 and 80\,M$_\odot$. Values between
2.1\,$\times$\,10$^8$\,M$_\odot$ and 3.6\,$\times$\,10$^8$\,M$_\odot$ are
obtained for the central $\sim$625\,pc region without including the young
stellar population responsible for the bright \HII\ region emission. {This
is consistent with our upper limit to the mass corresponding to the sum of the
dynamical mass found for the CNSFR and the nuclear clusters
(3.6\,$\times$\,10$^8$\,M$_\odot$).}

The values of dynamical masses of the individual clusters derived by us 
for the CNSFRs of both NGC\,2903 and NGC\,3351 (see Paper I) are larger than
those estimated for the individual clusters in the nuclear ring of NGC\,4314
from broad band U,B,V,I and H$\alpha$ photometry with HST
\citep{2002AJ....123.1411B} by more that an order of magnitude. 
{No comparable data exist for the CNSFRs studied here to allow the
estimate of photometric M/L ratios. However, the fact that the H$\beta$
luminosities of the NGC\,4314 clusters are lower than those of NGC\,2903 by a
factor of about 40 gives some support to our larger computed dynamical
masses. These masses are, on 
average, of the order of the largest kinematically derived mass for SSC in M82
\citep{2007ApJ...663..844M} from near IR spectroscopy
(40\,$\times$\,10$^5$\,M$_\odot$) but also larger (by factors between 4.2 and 34)
than the mass derived for the SSC A in NGC\,1569 by \cite{1996ApJ...466L..83H}
from stellar velocity dispersion measurements using red ($\sim$6000\,\AA)
spectra (3.3$\pm$0.5\,$\times$\,10$^5$\,M$_\odot$). 
}


{It should be recalled that} we have estimated the dynamical masses
through the virial theorem under the assumption  that the systems are
spherically symmetric, gravitationally bound and have isotropic velocity
distribution. We have used the stellar velocity dispersions derived from the
CaT absorptions features and the cluster sizes measured from the high spatial
resolution WFPC2-PC1 HST image. 

Therefore, while the average radius of an individual cluster is of the order of  2.8 pc, 
the stellar velocity dispersion corresponds to a much larger region, typically  
$\sim$\,42\,$\times$\,76\,pc$^2$ (1.0\,$\times$\,1.8\,arcsec$^2$) containing several knots. 
The use of these wider size scale velocity dispersion measurements to estimate the mass of each knot, can
lead to an overestimate of the mass of the individual clusters, and hence of each CNSFR (see Paper I).

However, as can be seen in the HST-NICMOS image (right-hand panel of
Fig.\ \ref{hst-slits}), the CNSFRs of NGC\,2903 show up very prominently in the near-IR
and dominate the light inside the apertures observed \cite[see][for a detailed
analysis of the IR images]{2001MNRAS.322..757A}. Therefore the assumption that the
light at the CaT wavelength region is dominated by the stars in the clusters seems to be justified.
The IR CaT is very strong, in fact the strongest stellar feature in very
young clusters, i.e. older than 4\,Myr \citep{1990MNRAS.242P..48T}.  Besides,
we detect a minimum in the velocity dispersion at the position of the
clusters (see Fig.\ \ref{dispersions-px}). Nevertheless, we can not assert that we are actually
measuring their velocity dispersion and thus prefer to say that our
measurements of $\sigma_{\ast}$, and hence dynamical masses, constitute upper
limits. Although we are well aware of the difficulties, yet we are confident
that these upper limits are valid and important for comparison with the gas
kinematic measurements.

{
Another important fact that can affect the estimated dynamical masses is
the presence of binaries among the red supergiant and red giant populations from
which we have derived the stellar velocity dispersions. 
In a recent work, Bosch et al.\ (2009) \nocite{2009AJ....137.3437B} using
GEMINI-GMOS data, have investigated the presence of binary 
stars within the ionizing cluster of 30~Doradus. From a seven epochs observing
campaign they have detected a rate of binary system candidates within their OB
star sample of $\sim$\,50\,\%. Interestingly enough,
this detection rate is consistent with a spectroscopic population of 100\,\%
binaries, when the observational parameters described in the simulations by
\cite{2001RMxAC..11...29B} are set for their observations. 
From their final sample of `single' stars 
they estimated a radial velocity dispersion of 8.3\,km\,s$^{-1}$. When they
derived $\sigma_{\ast}$ from a single 
epoch, they found values as high as 30\,km\,s$^{-1}$,
consistent with the values derived from  single epoch NTT observations by
\cite{2001A&A...380..137B}. 
}

Although the environment of our CNSFRs is very different from that of 30~Dor
and the stellar components of the binary systems studied by Bosch et al. (2009)
are very different from  the stars present 
in our regions from where the CaT arise (red supergiants), this is an
illustrative observational example of the problem. The orbital motions of
the stars in binary (multiple) systems produce an overestimate of the
velocity dispersions and hence of the dynamical masses. The single-star
assumption introduces a systematic error that depends on the properties of the
star cluster and the binary population, with an important effect on the
cluster mass if the typical orbital velocity of a binary component is of
the order of, or larger than, the velocity dispersion of the single/binary
stars in the potential of the cluster \citep{2008A&A...480..103K}. As was
pointed out by these authors, the
relative weights between the single and binary stars in the velocity dispersion
measurements depend on the binary fraction, which, together with the
semi-major axis or period distribution, are the most important parameters in order to
determine if the binary population affects  the estimated dynamical
masses. Their simulations indicate that the dynamical mass is overestimated by
70\,\% , 50\,\% and 5\,\% for a measured stellar velocity dispersion in the
line of sight of 1\,km\,s$^{-1}$, 2\,km\,s$^{-1}$ and 10\,km\,s$^{-1}$
respectively. They therefore conclude that most of the known dynamical masses 
of massive star clusters are only mildly affected by the presence of
binaries. Hence, although the binary
fraction of the red supergiants and red giants in this type of circumnuclear
and typically high metal rich environment is unknown, for our clusters, where
the smallest estimated velocity  
dispersion is 37\,km\,s$^{-1}$, we can assume that the contribution of
binaries to the stellar velocity dispersions is not important.

\section{Summary and conclusions}

We have measured gas and stellar velocity dispersions in four CNSFRs and
the nucleus of the barred spiral galaxy NGC\,2903. The stellar velocity dispersions
have been measured from the CaT lines at
$\lambda\lambda$\,8494, 8542, 8662\,\AA, while the gas velocity dispersions
have been measured by Gaussian fits to the H$\beta$\,$\lambda$\,4861\,\AA\ and
the [O{\sc iii}]\,$\lambda$\,5007\AA\ emission lines on high dispersion
spectra.

Stellar velocity dispersions are between 37 and 65\,km\,s$^{-1}$. These values
are about 25\,km\,s$^{-1}$ larger than those measured for the ionized gas from the
H$\beta$ emission line using a single Gaussian fit. On the other hand, single
Gaussian fits to the [O{\sc iii}]\,5007\,\AA\ line yield velocity dispersions
close to the stellar ones, and, in some cases, somewhat larger. However, the
best Gaussian fits involved two 
different components for the gas: a ``broad component" with a velocity
dispersion similar to that measured for the stars, and a ``narrow component"
with a velocity dispersion lower than the stellar 
one by about 30\,km\,s$^{-1}$. This last velocity component shows a relatively
constant value for the two gas emission lines, close 
to 24\,km\,s$^{-1}$ for all the studied CNSFRs (including the regions belonging to
NGC\,3351 from Paper I). We find a radial velocity shift
between the narrow and the broad component of the multi-Gaussian fits in the
CNSFRs of NGC\,2903 that varies between -10 and 35\,km\,s$^{-1}$.

When plotted in an [O{\sc iii}]/H$\beta$ versus [N{\sc ii}]/H$\alpha$
diagram, the two systems are clearly segregated for the high-metallicity
regions of NGC\,2903, with the narrow component showing lower excitation
and being among the lowest excitation line ratios detected within the SDSS
data set of starburst systems. Unfortunately, no comparable
information exists for the [N{\sc ii}]/H$\alpha$ ratio and hence we can not evaluate 
any possible contribution by shocks.

The {upper limits to the} dynamical masses estimated from the stellar
velocity dispersion using the 
virial theorem for the CNSFRs of NGC\,2903 are in the range between
6.4\,$\times\,10^7$ and 1.9\,$\times\,10^8$\,M$_\odot$ and is
1.1\,$\times\,10^7$\,M$_\odot$ for the nuclear region inside 3.8\,pc. Masses
derived from the H$\beta$ velocity dispersions under the assumption of a
single component for the gas would have been underestimated by factors between 
2 and 4 approximately. The {upper limits to the} derived masses for the
individual clusters are 
between 1.4\,$\times$\,10$^6$ and 1.1\,$\times$\,10$^7$\,M$_\odot$. These
values are between 4.2 and 34 times the mass derived for the SSC A in
NGC\,1569 by \cite{1996ApJ...466L..83H} and larger than other kinematically
derived SSC masses.

Masses of the ionizing stellar clusters of the CNSFRs have been derived from
their H$\alpha$ luminosities under the assumption that the regions are
ionization bound and without taking into account any photon absorption by
dust. For the star forming complexes of NGC\,2903 these masses are between 1.9 and
2.5\,$\times\,10^6$\,M$_\odot$, and is 2.1\,$\times\,10^5$\,M$_\odot$ for the galaxy
nucleus. Therefore, the ratio of the ionizing
stellar population to the total dynamical mass is between 0.01
and 0.04. These values of the masses of the ionizing stellar clusters of the
CNSFRs are comparable to that derived by \cite{1995ApJ...439..604G} for the
circumnuclear region A in NGC\,7714.

Masses for the ionized gas, also derived from their H$\alpha$ luminosities,
vary between 3.0\,$\times\,10^4$ and 7.9\,$\times\,10^5$\,M$_\odot$ for the
star forming regions and is 3\,$\times\,10^3$\,M$_\odot$ for the nucleus of
NGC\,2903. These values are also comparable to that derived by 
\cite{1995ApJ...439..604G}.

It is interesting to note that, according to our findings, the SSC in CNSFRs
seem to contain composite stellar populations. Although the youngest one
dominates the UV light and is responsible for the gas ionization, it
constitutes a few per cent of the total mass. This can explain the low
equivalent widths of emission lines measured in these regions.  This may well apply to
other SSC and therefore conclusions drawn from fits of single
stellar population (SSP) models should be taken with caution 
\citep[e.g.\ ][]{2003ApJ...596..240M,2004AJ....128.2295L}.
Furthermore, the composite
nature of the CNSFRs  means that star formation in the rings is a process that
has taken place over time periods much longer than those implied by the
properties of the ionized gas.

The observed stellar and [O{\sc iii}] radial velocities of NGC\,2903 are
in good agreement, while the H$\beta$ measurements show shifts similar to
those found between the narrow and the broad components. This different 
behaviour can be explained if the positions of the single Gaussian fits are
dominated by the broad component in the case of the [O{\sc iii}] emission line
and by the narrow one in the case of H$\beta$. The
rotation curve corresponding to the slit position through the nucleus shows
maximum and minimum values at the positions of the circumnuclear regions, as
observed in other galaxies with CNSFRs.

The existence of more than one velocity component in the ionized
gas corresponding to kinematically distinct systems, deserves further
study. Several results derived from the observations of the different emission
lines could be affected, among others: the classification of the activity in
the central regions of galaxies, the inferences about the nature of the source
of ionization, the gas abundance determinations, the number of ionizing
photons from a given region and any quantity derived from them. To
disentangle the origin of these two components it will be necessary to map
these regions with high spectral and spatial resolution and much better S/N
ratio in particular for the O$^{2+}$ lines. High resolution 3D spectroscopy
with IFUs would be the ideal tool to approach this issue.

\section*{Acknowledgements}
\label{Acknoledgement}


We acknowledge fruitful discussions with Guillermo Bosch, Nate Bastian and
Almudena Alonso-Herrero. 
We are grateful to Jay Gallagher for a thorough reading of the manuscript and
for suggestions that greatly improved its clarity.

The WHT is operated in the island of La Palma by the Isaac Newton Group
in the Spanish Observatorio del Roque de los Muchachos of the Instituto
de Astrof\'\i sica de Canarias. We thank the Spanish allocation committee
(CAT) for awarding observing time.

Some of the data presented in this paper and used in this work were obtained
from the 
Multimission Archive at the Space Telescope Science Institute (MAST). STScI is
operated by the Association of Universities for Research in Astronomy, Inc.,
under NASA contract NAS5-26555. Support for MAST for non-HST data is provided
by the NASA Office of Space Science via grant NAG5-7584 and by other grants
and contracts.

This research has also made use of the NASA/IPAC Extragalactic Database (NED)
which 
is operated by the Jet Propulsion Laboratory, California Institute of
Technology, under contract with the National Aeronautics and Space
Administration and of the SIMBAD database, operated at CDS,
Strasbourg, France. 

This work has been supported by Spanish DGICYT grants AYA-2004-02860-C03 and
AYA2007-67965-C03-03. GH and MC 
acknowledge support from the Spanish MEC through FPU grants AP2003-1821 and
AP2004-0977. AID acknowledges support from  the Spanish MEC through a
sabbatical grant PR2006-0049. Furthermore, partial support from the Comunidad
de Madrid under grant S-0505/ESP/000237 (ASTROCAM) is acknowledged. Support
from the Mexican Research Council (CONACYT) through grant 19847-F is
acknowledged by ET and RT. We thank the hospitality of the Institute of
Astronomy, Cambridge, where part of this work was developed. GH and MC also
thank the hospitality of the INAOE, and ET and RT, that of the Universidad
Aut\'onoma de Madrid.

\bibliographystyle{mn2e}
\bibliography{ngc2903}

\end{document}